\documentclass[iop,apj,twocolappendix,numberedappendix]{emulateapj}

\usepackage{rotating}

\usepackage{float}
\floatstyle{plain}
\newfloat{sidefig}{thp}{lop}
\floatname{sidefig}{}

\newcommand{\degree}{^o}

\newcommand{\KeV}{\,\textrm{keV}}

\newcommand{\Kpc}{\,\textrm{kpc}}

\newcommand{\Kms}{\,\textrm{km}\,\textrm{s}^{-1}}

\newcommand{\ccm}{\,\mathrm{cm}^{-3}}
\newcommand{\gccm}{\,\mathrm{g}\,\mathrm{cm}^{-3}}

\newcommand{\Myr}{\,\mathrm{Myr}}

\newcommand{\Reyn}{\textrm{Re}}

\shorttitle{Stripping of M89}
\shortauthors{Roediger et al.}

\begin{document}


\title{Stripped elliptical galaxies as probes of ICM physics: II. Stirred, but mixed? 
Viscous and inviscid gas stripping of the Virgo elliptical M89}


\author{E.~Roediger\altaffilmark{1,2,5},  R.~P.~Kraft\altaffilmark{2}, P.~E.~J.~Nulsen\altaffilmark{2}, W.~R.~Forman\altaffilmark{2}, M. Machacek\altaffilmark{2}, S. Randall\altaffilmark{2}, C.~Jones\altaffilmark{2},  E.~Churazov\altaffilmark{3}, R.~Kokotanekova\altaffilmark{4}}
\affil{
\altaffilmark{1}Hamburger Sternwarte, Universit\"{a}t Hamburg, Gojensbergsweg 112, D-21029 Hamburg, Germany\\
\altaffilmark{2}Harvard/Smithsonian Center for Astrophysics, 60 Garden Street MS-4, Cambridge, MA 02138, USA\\
\altaffilmark{3}MPI f\"{u}r Astrophysik, Karl-Schwarzschild-Str. 1, Garching 85741, Germany\\
\altaffilmark{4}AstroMundus Master Programme, University of Innsbruck, Technikerstr. 25/8, 6020 Innsbruck, Austria
}

\email{eroediger@hs.uni-hamburg.de}

%
%
\altaffiltext{5}{Visiting Scientist, SAO}

\begin{abstract}
Elliptical  galaxies  moving through the intra-cluster medium (ICM) are {progressively} stripped of their gaseous atmospheres.  X-ray observations reveal the structure of galactic tails, wakes, and the interface between the galactic gas and the ICM. This fine-structure depends on dynamic conditions (galaxy potential, initial gas contents, orbit in the host cluster), orbital stage (early infall, pre-/post-pericenter passage),  as well as on the still ill-constrained ICM plasma properties (thermal conductivity, viscosity, magnetic field structure). 
Paper I describes flow patterns and stages of inviscid gas stripping. Here we study the effect of a Spitzer-like temperature dependent viscosity corresponding to Reynolds numbers, Re,  of 50 to 5000 with respect to the ICM flow around the remnant atmosphere. Global flow patterns are independent of viscosity in this Reynolds number range. 
Viscosity influences two aspects:  In inviscid stripping, Kelvin-Helmholtz instabilities (KHIs) at the sides of the remnant atmosphere lead to observable horns or wings. Increasing viscosity suppresses KHIs of increasing length scale, and thus observable horns and wings.  Furthermore, in inviscid stripping, stripped galactic gas can mix with the ambient ICM in the galaxy's wake. This mixing is suppressed increasingly with increasing viscosity, such that viscously stripped galaxies  have  long X-ray bright, cool wakes. We provide mock X-ray images for different stripping stages and conditions. While these qualitative results are generic, we tailor our simulations to the Virgo galaxy M89 (NGC 4552), where  $\Reyn\approx 50$ corresponds to a viscosity of 10\% of the Spitzer level.  Paper III compares new deep Chandra and archival XMM-Newton data  to our simulations.   
\end{abstract}


\section{Introduction} \label{sec:intro}

Elliptical galaxies falling into clusters experience an  intra-cluster medium (ICM) head wind that   progressively strips their gaseous atmospheres (\citealt{Nulsen1982,Gisler1976,Takeda1984,Stevens1999,Toniazzo2001,Acreman2003,McCarthy2008}, among others). Previous work shows that the stripped gas forms a `tail' downstream of the galaxy.   In a companion paper we clarify the nature  of  galaxies' gas `tails' and distinguish between the galaxy's \textit{remnant tail} and its \textit{wake} (Roediger et al., paper I hereafter). 
We showed that the downstream part of an atmosphere undergoing gas stripping is shielded from the ICM head wind and can be largely retained by the galaxy up to or beyond pericenter passage. {The retained downstream atmosphere can take the appearance of a tail. This `remnant tail' should not be confused with the `tail of stripped gas' because it is not stripped but part of the remnant atmosphere.} In analogy to the flow around a solid body, the ICM flow \textit{around} the remnant atmosphere leads to a wake  downstream of the remnant atmosphere. The wake is filled with both stripped galactic gas and ICM. The near part of this wake is a `deadwater' region which extends roughly one or two  atmosphere lengths. In X-ray observations of stripped ellipticals, both the remnant tail and the near wake can have a tail-like appearance.

\begin{figure*}
\includegraphics[trim= 0 0 0 0,clip,angle=0,width=0.99\textwidth]{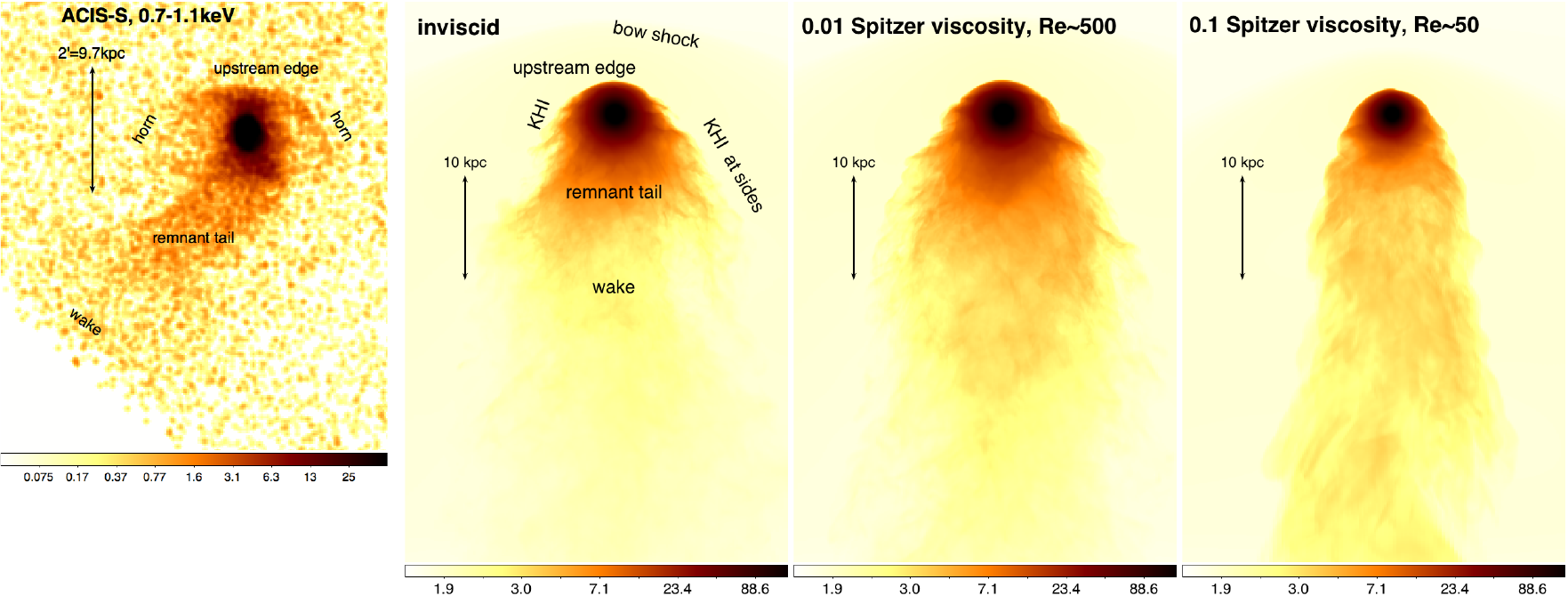}
 \caption{\textbf{Left}: Merged Chandra ACIS-S image of gas-stripped elliptical galaxy M89 in Virgo, showing the truncated galactic atmosphere with an upstream edge in the north, a downstream tail and the onset of the wake towards the south, and horns at the sides (200 ks total, 0.7-1.1 keV band, point sources removed, binned to 2 arcsec pixels and smoothed with a Gaussian kernel with 2 pixel radius). We label  structures discussed in this paper.  
\textbf{Second panel}: Synthetic X-ray image from our best-matching simulation  in the same energy band, with structures labelled (inviscid stripping, initially extended atmosphere, 140 Myr after pericenter passage, orbit inclined by 30 degree out of the plane of the sky; same as third panel in Fig.~\ref{fig:Xray_inclination} but flipped around the vertical axis; brightness units are arbitrary). The simulation reproduces the bright near tail and faint far tail or wake, and the upstream edge. M89's horns resemble the KHIs at the sides of the simulated galaxy but are unusually close to the upstream edge given their size.
\textbf{Third panel}: Corresponding viscous simulation with 0.01 Spitzer viscosity.  Small-scale KHIs are suppressed, thus the wake remains brighter to somewhat farther downstream.
\textbf{Fourth panel}:  Corresponding viscous simulation with 0.1 Spitzer viscosity.  KHIs up to scales comparable to the radius of the remnant atmosphere are suppressed, leading to suppression of mixing in the wake and consequently a bright `tail' of 10s of kpc length.}
\label{fig:m89_plain}
\end{figure*}


The Chandra and XMM-Newton X-ray observatories observed gas stripping for several nearby elliptical galaxies. The left panel of Fig.~\ref{fig:m89_plain} shows a  deep Chandra image of the strongly gas-stripped  elliptical M89 (NGC 4552), which is located 350 kpc (72 arcmin) east of the Virgo cluster center  (M87). Like other stripped ellipticals, it shows a truncated atmosphere with a gas `tail' and a contact discontinuity on opposite sides, indicating, in this case, a northward motion through the Virgo ICM. Beyond these basic features, this galaxy displays two `horns' that are attached to its upstream edge and bend downstream (\citealt{Machacek2006a}). Its near  `tail' curves to the east, the far `tail' appears to flare near the edge of the field of view. Observations of other stripped ellipticals display similarly rich structures: M86 in Virgo shows a spectacular 150 kpc long bifurcating `tail' starting in a plume (\citealt{Randall2008}), M49 in Virgo (\citealt{Kraft2011}) has a ragged upstream edge and a flaring `tail', and the `tail' of NGC 1404 in Fornax appears rather faint despite its projected proximity to the cluster center (\citealt{Machacek2005}).

The  details and differences of the gas-stripped cluster ellipticals could be used to determine the still ill-constrained effective ICM transport coefficients.   Temperature discontinuities across their upstream edges suggest a strongly suppressed thermal conductivity (e.g., \citealt{Vikhlinin2001,Vikhlinin2002}).  The flow  around the remnant atmosphere of a galaxy is also shaped by the effective ICM viscosity and/or magnetic field structures in the ICM (\citealt{Dursi2008}). Here we focus on the impact of an {isotropic} effective ICM viscosity. 

The patterns of viscous and inviscid flows around  solid bodies are well-known. Recall, e.g., the flow  past a circular cylinder or sphere (\citealt{vanDyke1982}, among others): 
From high to low viscosity, or from low to high Reynolds number%
\footnote{The Reynolds number is defined as $\Reyn=\frac{l v \rho}{\mu}$ where $l$ and $v$ are the characteristic length scale and velocity of the flow, and $\rho$ and $\mu$ are the fluid density and dynamic viscosity. Reynolds numbers stated in this paper refer to diameter of the remnant atmosphere, the galaxy's orbital velocity, ambient ICM density and its viscosity. }%
, flow patterns change from laminar to a downstream vortex pair or torus to a turbulent wake (see also Fig.~\ref{fig:boxflow} in the appendix). The wake of a stripped galactic atmosphere behaves similarly, and stripped gas traces the flow patterns in the wake. In a turbulent wake the stripped gas should mix quickly with the ambient ICM, reducing the gas density and hence X-ray brightness. {Increasing viscosity suppresses Kelvin-Helmholtz instabilities (KHIs) at larger and larger scales (\citealt{Roediger2013khi}) and thus suppresses turbulent mixing in the galaxy's wake.} Therefore, for a sufficiently viscous ICM, we expect an  unmixed, cool, X-ray bright wake.  Mock X-ray images of our simulations in  Fig.~\ref{fig:m89_plain} show these features in direct comparison with the M89 observation.
 
The mixing state, and thus brightness of observed tails and wakes of stripped ellipticals does not only depend on the ICM transport properties but also on dynamical conditions such as galaxy potential, initial galactic atmosphere, and the ICM flow experienced along the orbit through the cluster.  To facilitate the most direct comparison to observations, the simulations presented here are tailored to M89 regarding these dynamical conditions as described in paper I. Qualitative results are independent of this particular model galaxy. 

Here we investigate the impact of an isotropic, Spitzer-like\footnote{dynamic viscosity is temperature-dependent, $\propto T^{5/2}$, \citet{Spitzer1956}} viscosity on the flow patterns in the stripping process and discuss observable morphological differences between viscous and inviscid stripping.  
We do so by simulating an elliptical, or rather spherical, galaxy during cluster crossing with viscous hydrodynamics as described in Sect.~\ref{sec:sim_setup}. In Sect.~\ref{sec:viscflowpatterns} we compare viscous and inviscid flow patterns and their evolution. Sect.~\ref{sec:observables} discusses the appearance of these patterns in X-ray images. In Sect.~\ref{sec:discussion} we briefly compare our results to M89 and to other stripped ellipticals, and discuss additional physics. A detailed comparison to the new M89 Chandra data is presented in a companion paper (paper III, Kraft et al. in prep.). Section~\ref{sec:summary} summarizes our results.

%
%
\section{Method}  \label{sec:sim_setup}
We model the motion of a spherical galaxy through its host cluster with 3D hydrodynamic simulations. The simulations are run in the rest frame of the galaxy, i.e., the galaxy is exposed to an ICM head wind which varies in density and velocity according to the galaxy's orbit though the host cluster.  The model galaxy consists of an analytic gravitational potential due to its dark matter and stellar contents, and an initially hydrostatic $\sim 0.4\KeV$ hot  atmosphere.  Tailoring the galaxy model and  ICM wind to M89 is described in Appendix A in paper I. In paper I we estimated that gas replenishment by stellar mass loss is not relevant in our simulations and is thus neglected.

We model both the galactic atmosphere and the ICM as viscous or inviscid gases with an ideal equation of state. Thermal conduction is neglected. We do not include radiative cooling or AGN heating of the galactic gas but assume that thermal balance is maintained by the interplay between both.

\subsection{Code}

We use the FLASH code (version 4.0.1, \citealt{Dubey2009}),   a modular, block-structured AMR code, parallelized using the Message Passing Interface library. It solves the Riemann problem on a Cartesian grid using the Piecewise-Parabolic Method. Appendix~\ref{sec:solidbodysims} briefly describes test simulations of inviscid and viscous flows around a solid body.

{The simulations use a nested grid centered on the galaxy and its near wake and ensure the peak resolution in a box of at least $(-10\Kpc, 50\Kpc)\times (\pm 10\Kpc)^2$ around the galaxy center. At early times the upstream and side extents of the peak resolution box are larger to encompass the whole atmosphere. As stripping proceeds, the upstream extent of the peak resolution box  is reduced to $\sim 2\Kpc$ ahead of the remnant atmosphere to avoid unnecessary refinement of the free-streaming upstream ICM. For the viscous simulations a peak resolution of 0.2 kpc is sufficient to resolve all KHIs that are not suppressed by viscosity. This lower peak resolution compared to the inviscid simulations (which use 0.1 kpc, see paper I) also reduces computational costs for the more expensive viscous simulations.}
  We confirmed that the resolution is sufficient for convergence of our results (App.~\ref{sec:resolution}).

{Both the inviscid and viscous} simulations include spatial perturbations in the ICM head wind as seeds for KHIs   as described in Paper I.

\subsection{Viscosity}

We adapted the implicit thermal diffusion module of FLASH to implicit momentum diffusion to describe viscous effects. Viscous heating is included as a separate energy source term. We use an isotropic,  temperature dependent Spitzer-like viscosity with a dynamic viscosity $\mu \propto T^{5/2}$ (\citealt{Spitzer1956}). For our viscous simulations, we use a dynamic viscosity of  0.001 to 0.1 of the Spitzer value for an unmagnetized plasma. For an ICM temperature of 2.4 keV, electron density $3\times 10^{-4}\ccm$,  atmosphere diameter of 6 kpc, and  ICM wind velocity of $\sim 1000\Kms$ a viscosity of  0.1 of the Spitzer value corresponds to a Reynolds number, Re,  of 46. Using instead the shocked ICM temperature near the stagnation point of $\sim 4.5$ keV results  Re $\sim 10$. Despite being only 10\% of the Spitzer-value, this viscosity pushes the hydrodynamic approximation to the limit because the corresponding mean free path is 0.6 kpc, a considerable fraction of the size of the remnant atmosphere.

\subsection{Simulation runs}
{We repeat the same two simulations as in paper I, i.e., stripping of an initially compact and an initially extended atmosphere, with 0.1 Spitzer viscosity (Re at pericenter $\sim$ 46). We also simulate stripping of the extended atmosphere at 0.01 and 0.001 Spitzer viscosity (Re at pericenter $\sim$ 460 and 4600), as well as with a spatially constant kinematic viscosity.}

Galaxy models and orbits are described in Appendix A in paper I.

\section{Comparison of inviscid and viscid flow patterns and their evolution} 
\label{sec:viscflowpatterns}
%
Figures~\ref{fig:dens_compactSpitzer} and \ref{fig:dens_extended_Spitzer} show the evolution of 0.1-Spitzer viscous stripping of the  compact and extended atmosphere, respectively. They can be compared directly to the inviscid cases shown in Figs.~2 and 9 in paper I.  Figure~3 in paper I  provides a side-by-side comparison of the flow patterns for all simulation runs (inviscid/viscous stripping, compact/extended atmosphere).

\subsection{Similar global flow patterns in inviscid/viscous stripping}
%

\begin{figure*}
\includegraphics[angle=0,width=0.99\textwidth]{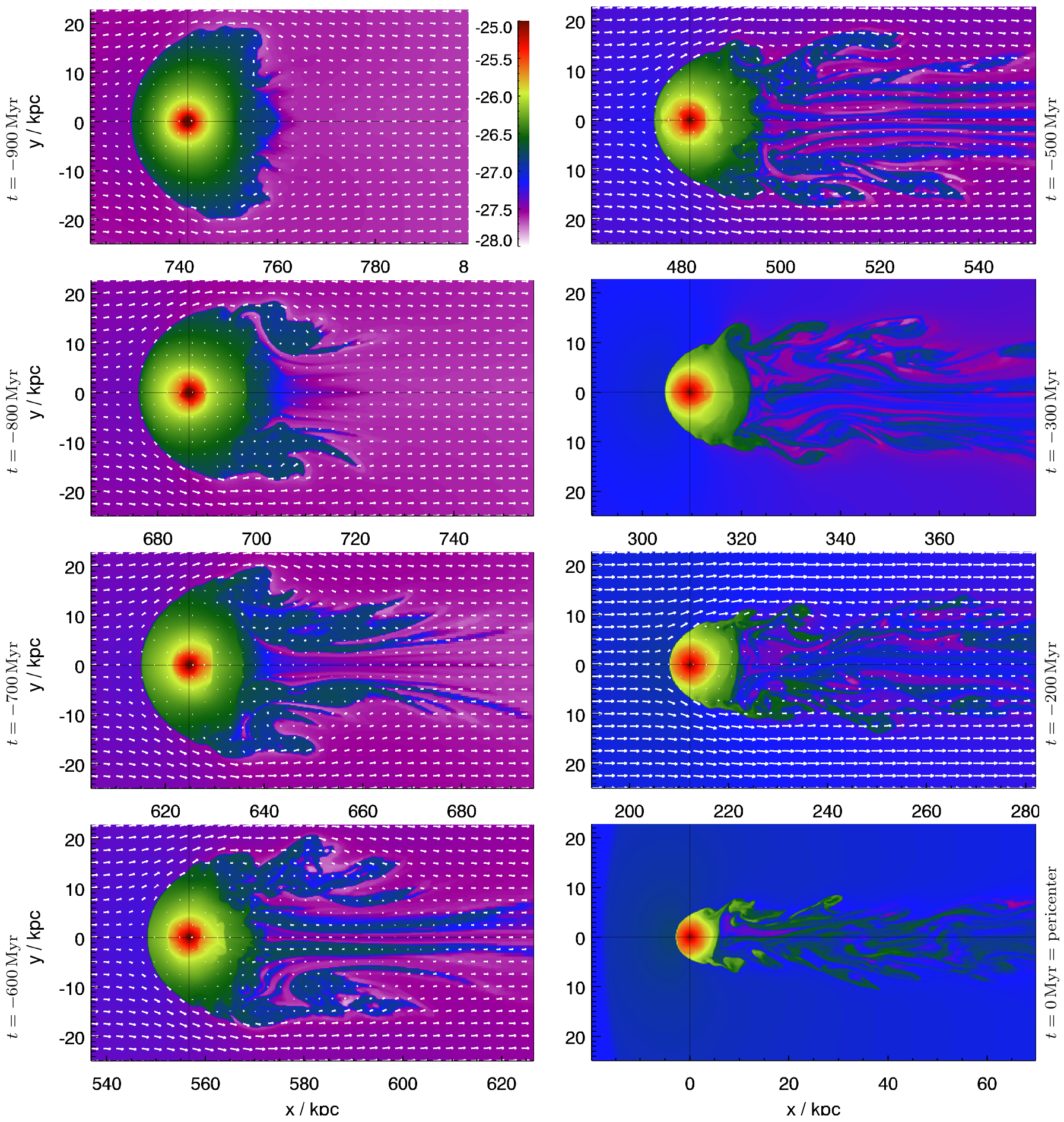}
\caption{Stripping of a compact galactic atmosphere with 0.1 Spitzer viscosity. (Sect.~\ref{sec:viscflowpatterns}). 
Density slices through galaxy in orbital plane, overlaid with velocity vectors in most panels. The left column shows the initial relaxation phase, the right column the quasi-equilibrium phase. In contrast to inviscid stripping (Fig.~2 in paper I), KHIs at the sides of the galaxy are mostly suppressed. Furthermore, the suppressed mixing in the wake leads to cool dense filaments in the wake. 
\label{fig:dens_compactSpitzer}}
\end{figure*}


\begin{figure*}
\includegraphics[angle=0,width=0.99\textwidth]{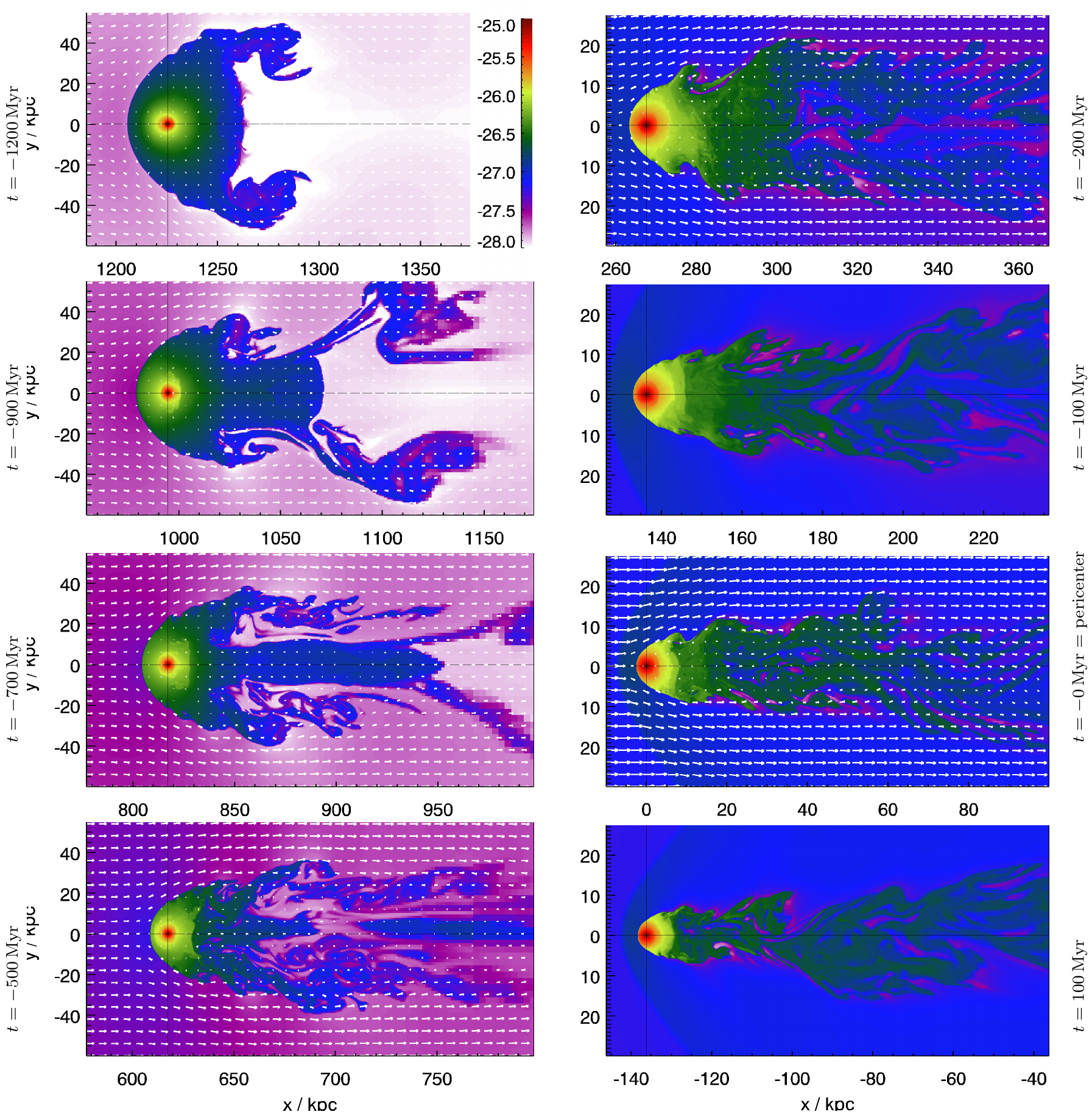}
\caption{Stripping of initially extended atmosphere with 0.1 Spitzer viscosity  (Sect.~\ref{sec:viscflowpatterns}).  
Density slices through galaxy in orbital plane with velocity vectors overlaid in most panels. The left column shows the initial relaxation phase, the right column the quasi-equilibrium flow. Compare to Fig.~9 in paper I for the corresponding inviscid stripping.
\label{fig:dens_extended_Spitzer}}
\end{figure*}


The global evolution of the remnant atmosphere as well as the global  patterns in the ICM flow around   it are independent of viscosity in the range covered here. In both inviscid and viscous stripping, the quasi-steady flow stage is reached only after an extended relaxation phase that follows the onset of the ICM flow (left columns of Figs~\ref{fig:dens_compactSpitzer} and \ref{fig:dens_extended_Spitzer}). As explained in Paper I (Section 3.2.1), the primary characteristic of the relaxation phase is a series of two or three oscillations of the downstream atmosphere along the direction of motion before a proper wake is established.  Viscosity somewhat damps these oscillations but does not erase them. 

Independent of viscosity, during quasi-steady stripping gas removal from a stripped atmosphere occurs predominantly along the sides of the atmosphere. The downstream atmosphere is shielded from the ICM wind and can survive until after pericenter passage, forming the galaxy's remnant tail (right columns of Figs.~\ref{fig:dens_compactSpitzer} and \ref{fig:dens_extended_Spitzer}, compare to Figs.~2 and 9 in paper I).   

Independent of viscosity the wakes do not flare dramatically but remain narrow as seen in wakes of transonic solid bodies. The wakes also have rather sharp, albeit irregular, boundaries to the ambient ICM. In both inviscid and viscous stripping, the galaxy's near wake is characterized by a deadwater region. The length of the remnant tail and  the contents of the deadwater region depend on the initial atmosphere as shown in paper I. The mixing in the deadwater region and in the far wake depend on the ICM viscosity as discussed below. 

The global similarities are partly due to the strong temperature dependence of the Spitzer-like viscosity, which leads only to a very mild viscosity in the cooler galactic gas despite the substantial viscosity in the hotter ICM.

\subsection{Differences with viscosity}

The crucial differences between inviscid and viscous stripping lie in the gas removal mechanism and in the mixing in the galaxy's wake. 

\subsubsection{Viscous gas removal mechanism}

In inviscid stripping, the momentum transfer from the ICM into the galactic gas at the sides of the atmosphere occurs via KHIs. The KHI rolls start roughly 45 degrees away from upstream stagnation point,  grow while being driven  along the sides of the atmosphere,  are sheared off by the ICM wind and  mixed into the ICM. For the initially compact atmosphere, this  shapes the  tail of stripped gas into a hollow cylinder whose walls are made of filaments of stripped galactic gas (Fig.~2 in paper I). For the initially extended atmosphere, the KHI rolls are driven along the full length of the remnant atmosphere, including the remnant tail, where they grow to a size larger than the upstream or side radius of the atmosphere (Fig.~9 in paper I). 

In viscous stripping at 0.1 Spitzer viscosity,  momentum is transferred from the ICM into the galactic gas  directly via viscosity (Figs.~\ref{fig:dens_compactSpitzer} and \ref{fig:dens_extended_Spitzer}). For this viscosity, KHIs can start to grow, at reduced speed, only for scales larger than  $\sim 6\Kpc$, but die down again, in agreement with \citet{Roediger2013khi}. Even larger KHI rolls die down, and KHIs on smaller scales are fully suppressed. 

Stripping at 0.001 Spitzer viscosity is almost indistinguishable from inviscid stripping. At 0.01 Spitzer viscosity, KHIs below $\sim 2$ kpc can grow only at reduced speed, and die down quickly, but KHIs on scales of $5$ kpc and above can grow. As the dominant KHIs mode are the largest ones, i.e., comparable to the remnant atmosphere radius,  the gas removal occurs still via KHIs. The absence of the smaller KHI modes is only a subtle effect.

\subsubsection{Reduced mixing in the wake}

 In the same manner viscosity {reduces} turbulence and mixing in the wake. At 0.1 Spitzer viscosity, cold stripped gas forms long filaments in the tail, similar to the filaments that occur when stirring two viscous paints (Figs.~\ref{fig:dens_compactSpitzer} and \ref{fig:dens_extended_Spitzer}). Consequently, in the viscous wake, filaments of cool gas and hot ICM co-exist for tens of kpc downstream of the galaxy.  {In the cold filaments the galactic gas fraction remains higher than 80\% beyond 100 kpc from the galaxy.}   Thus, filamentary tails can be produced by viscous stripping, without the presence of magnetic fields. It will be interesting to try to distinguish both scenarios. 
 
 In  inviscid stripping  the global flow patterns in the wake are superimposed with irregular velocity fluctuations of a few $100\Kms$ (see Fig.~5 in paper I) {that mix ambient ICM into the wake as shown by the decreasing galactic gas fraction in the wake (Fig.~4 in paper I). } At  intermediate viscosity of 0.01 Spitzer, mixing along the wake is somewhat slower, leading to a somewhat cooler wake temperature at a given distance down the wake (Fig.~4 in paper I). 
 
\begin{figure}
\centering\includegraphics[trim= 0 0 0 0,clip,angle=0,width=0.5\textwidth]{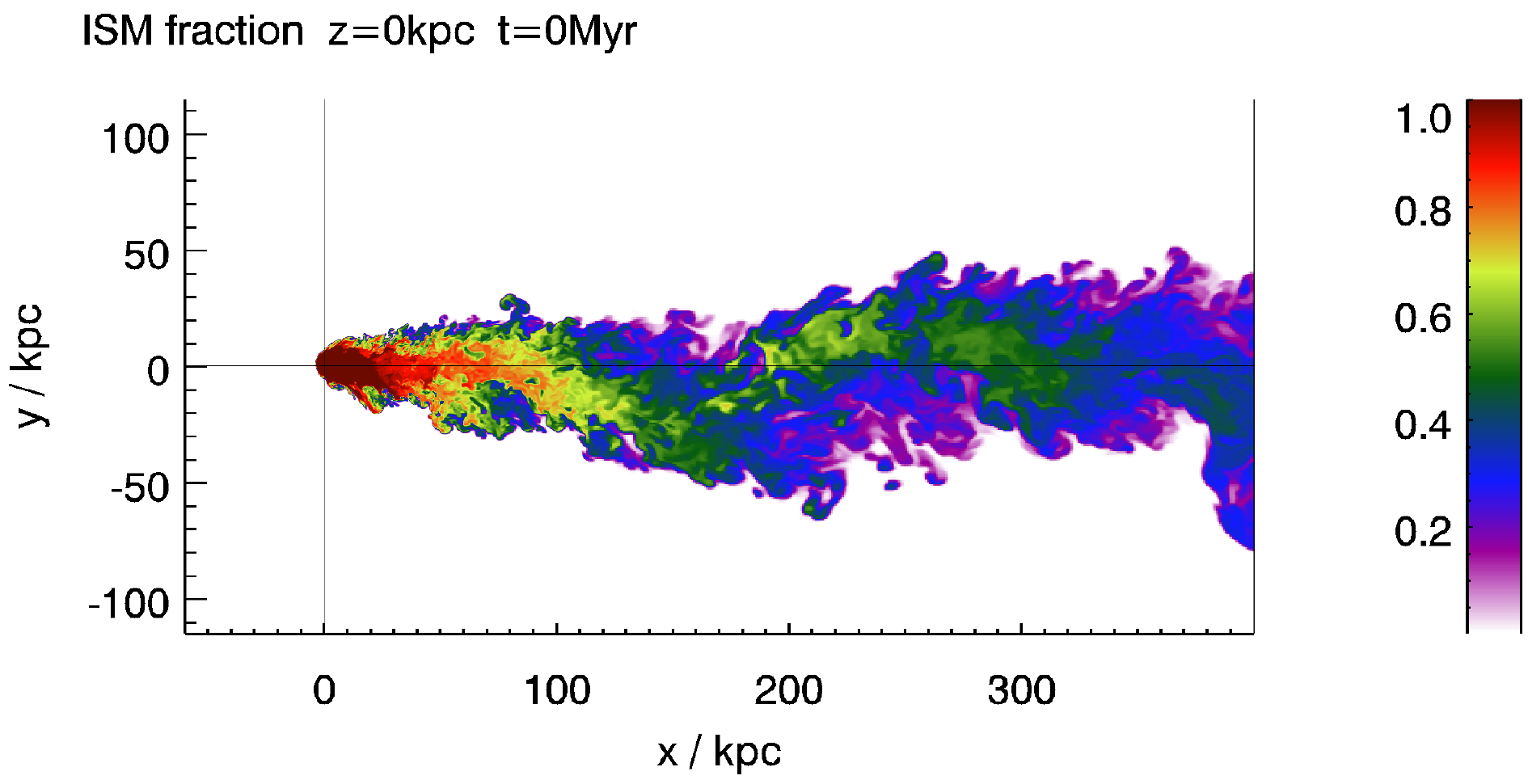}
\caption{Fraction of galactic gas in the wake of the in viscidly stripped galaxy (initially extended atmosphere) at pericenter passage. The wake is a coherent structure hundreds of kpc downstream of the galaxy, although the galaxy has already reached pericenter passage. The same effect is observed in the wakes of supersonic bullets in the earth's atmosphere, where the wake remains coherent out to several hundred bullet diameters downstream of the bullet (Fig.~151 in \citealt{vanDyke1982}).
\label{fig:longwake}}
\end{figure}

 {We note, however, that the galaxy's wake remains a coherent structure for hundreds of kpc downstream of the galaxy even in turbulent stripping (Fig.~\ref{fig:longwake}), an effect known from bullets in the earth's atmosphere (Fig.~151 in \citealt{vanDyke1982}). The average galactic gas fraction decreases along and perpendicular to the wake, where the gradient depends on viscosity and also strongly on initial gas contents of the galaxy (Fig.~4 in paper I). }

\subsection{Impact of viscous heating}

{We also performed test runs that neglected viscous heating and found }no significant effect on the flow patterns described above. However, {the presence of viscous heating creates} a hot layer in the ICM at the upstream edge and along the sides of the atmosphere and the wake (Fig.~4 in paper I). Consequently, due to pressure balance, the gas density in this layer is reduced, and the atmosphere and wake are surrounded by a depletion layer. Our simulations may overestimate viscous heating because we did not take into account possible saturation of viscosity. However, qualitatively the effect should exist. Magnetic field draping is expected to lead to a similar depletion layer due to the magnetic pressure in the draping layer. Careful considerations are required to distinguish both origins of the depletion layer observationally.

\subsection{Impact of spatially constant viscosity} \label{sec:viscousflow_allvisc}
%

\begin{figure}
\includegraphics[angle=0,width=0.47\textwidth]{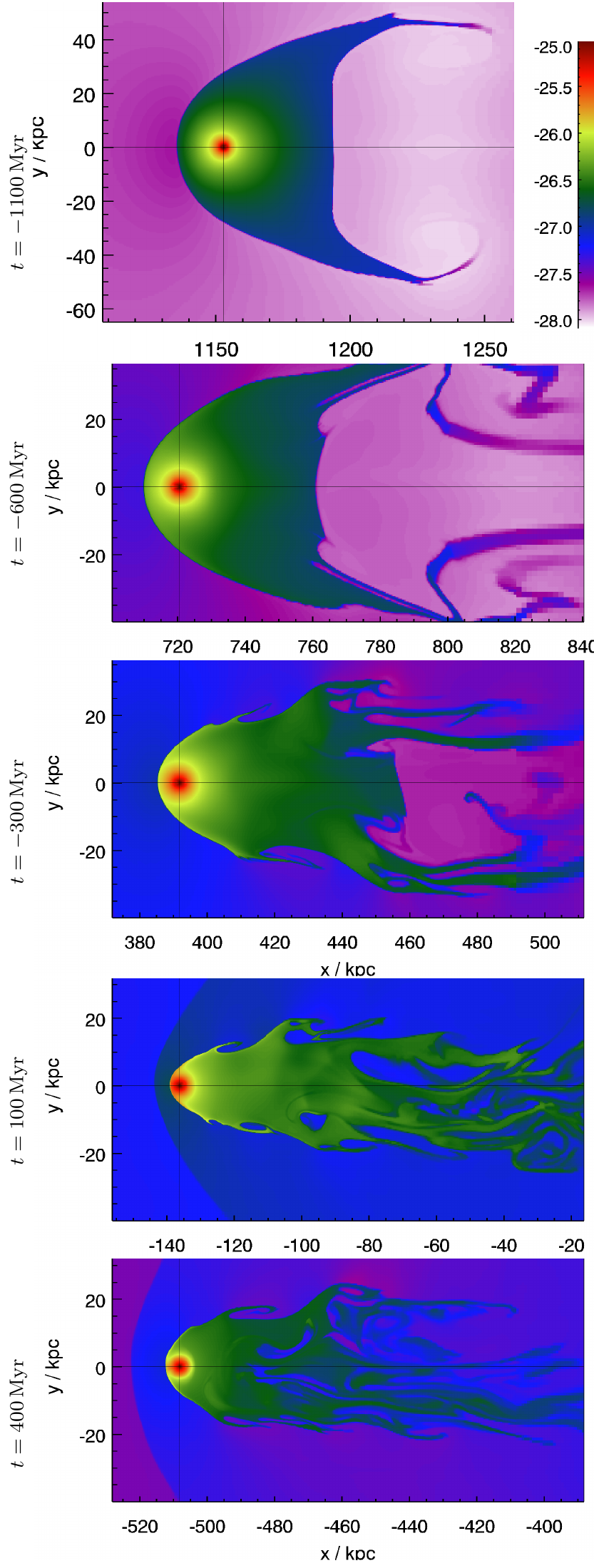}
\caption{Stripping of initially extended atmosphere with spatially constant kinematic viscosity, density slices through the orbital plane (Sect.~\ref{sec:viscousflow_allvisc}). The atmosphere is not truly stripped but deformed and shows a pronounced remnant tail.
\label{fig:dens_extended_constvisc}}
\end{figure}


 For academic interest we ran a simulation with a constant kinematic viscosity (Fig.~\ref{fig:dens_extended_constvisc}) of $\nu=10^{28}$ cm$^2$ s$^{-1}$ throughout the simulation box (no viscous heating), corresponding to a few percent of the Spitzer level in the ICM, but exceeding the Spitzer level in the cooler galactic gas by a factor of 50 to 500 for gas densities of $10^{-2}$ to $10^{-1}\ccm$, respectively (mean free path in the galactic gas is 0.4 kpc). The high viscosity of the galactic gas takes the shielding of the downstream atmosphere to the extreme. {Gas is hardly torn off the atmosphere, but the flow around the atmosphere merely distorts it. The ambient flow stretches the atmosphere  strongly into the downstream direction, where the galactic atmosphere settles into a new equilibrium and is again shielded from the ICM head wind and kept in place by the galaxy's gravity.}  
 The effect is most impressive for the initially extended atmosphere (see Fig.~\ref{fig:dens_extended_constvisc} for a time series, and Fig.~4 in paper I for comparison to other runs). As now also the galactic gas is viscous, KHIs and mixing are suppressed up to scales of $\sim 15\Kpc$. As a result, viscous momentum transport at the sides of the atmosphere can pull the outer layers in the downstream direction, which leads to the stretching of the atmosphere, but can hardly tear off  filaments.   Even at  pericenter passage, most of the galactic gas still resides in the remnant tail that stretches 60 kpc behind the galaxy center, although the upstream radius is only 3 kpc as in the other simulations.  Thick filaments are sheared off only beyond 60 kpc downstream of the galaxy center. The galaxy still has an extremely long remnant tail at 0.5 Gyr after pericenter passage, where the ICM wind  decreases and the atmosphere starts to re-settle.

\section{Characteristics of viscosity and dynamic conditions in mock X-ray images} 
\label{sec:observables}
%

We calculate  synthetic X-ray images in the 0.7-1.1 keV band, i.e., around the Fe L emission line complex, by projecting $n^2 \Lambda(T)$ perpendicular to the orbit. The band is chosen to highlight the cooler galactic gas ($\sim 0.5\KeV$) over the hotter ICM ($\gtrsim 2.4\KeV$). The cooling function $\Lambda$ is calculated with XSPEC assuming the APEC model.  We do not include count noise or smoothing by a specific instrumental point spread function to enable easier comparison to different instruments and observations. The projections  contain  a LOS through our simulation box around the orbital plane of the galaxy such that  all stripped gas and  little of the ambient ICM is included in the projection. We add the cluster background as described below. Figure~\ref{fig:m89_plain} displays the corresponding Chandra image of M89. 

For snapshots from the quasi-steady stripping phase we add the emission of the Virgo cluster ICM  at the true projected position of M89 in Virgo, i.e., at 350 kpc from the cluster center, independent of the current position of the simulated galaxy in the cluster. By doing so, we assume that the orbital plane of M89 is inclined away from the plane of the sky sufficiently to give a projected distance to the cluster center as observed for M89. For snapshots from the initial relaxation phase, we add the  cluster background at the galaxy's current projected position  assuming that we see the galaxy moving in the plane of the sky.

\subsection{Initial relaxation phase}  \label{sec:obs_relax}

\begin{figure*}
\textbf{initially extended atmosphere, inviscid stripping, initial relaxation phase}\\
\centering\includegraphics[trim= 0 0 0 0,clip,angle=0,width=0.93\textwidth]{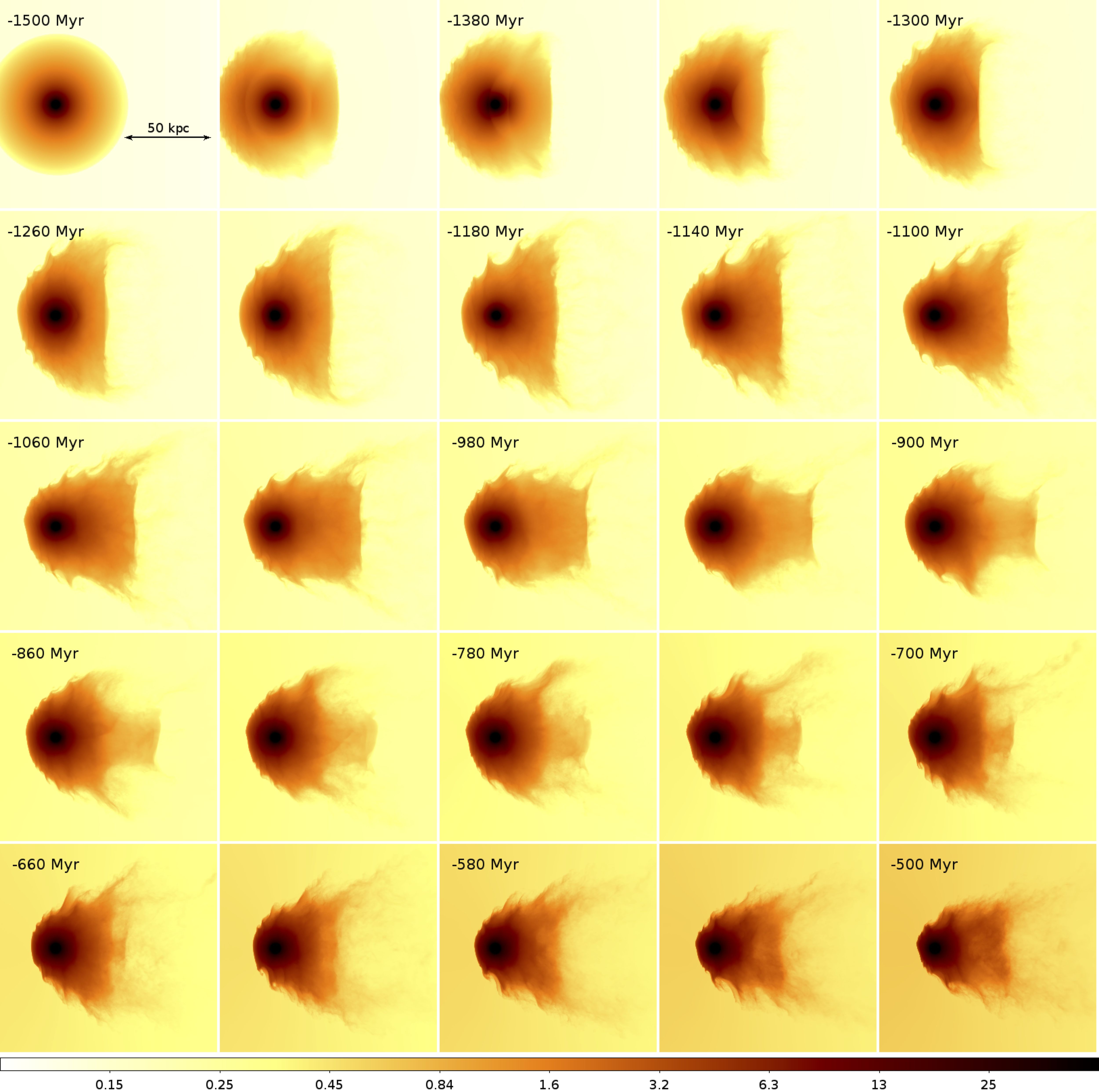}
\caption{X-ray images (0.7-1.1 keV) of simulated gas-stripped galaxy, surface brightness in arbitrary units. Initially extended atmosphere, inviscid stripping. Initial relaxation phase, timesteps of 40 Myr, pericenter passage occurs at $t=0$. See Sect.~\ref{sec:obs_relax}. 
\label{fig:Xray_ini}}
\end{figure*}
\nopagebreak
\begin{figure*}
\textbf{initially extended atmosphere, inviscid stripping, initial relaxation phase}\\
\centering\includegraphics[trim= 0 0 0 0,clip,angle=0,width=0.93\textwidth]{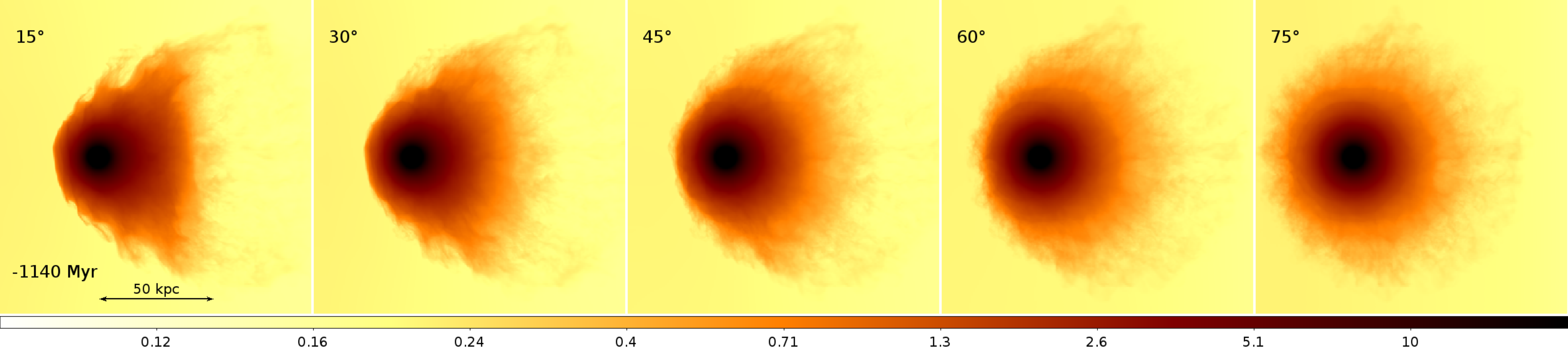}
\caption{Impact of inclination on inviscidly stripped atmosphere during initial relaxation phase for timestep $t=-1140 \Myr$ (cf.~Fig.~\ref{fig:Xray_ini}, second row, fourth column). The galaxy is rotated around the vertical axis by the indicated angle.
\label{fig:Xray_ini_rot}}
\end{figure*}

\begin{figure*}
\centering\textbf{initially extended atmosphere, 0.1 Spitzer viscosity stripping, initial relaxation phase}\\
\includegraphics[trim= 0 0 0 0,clip,angle=0,width=0.93\textwidth]{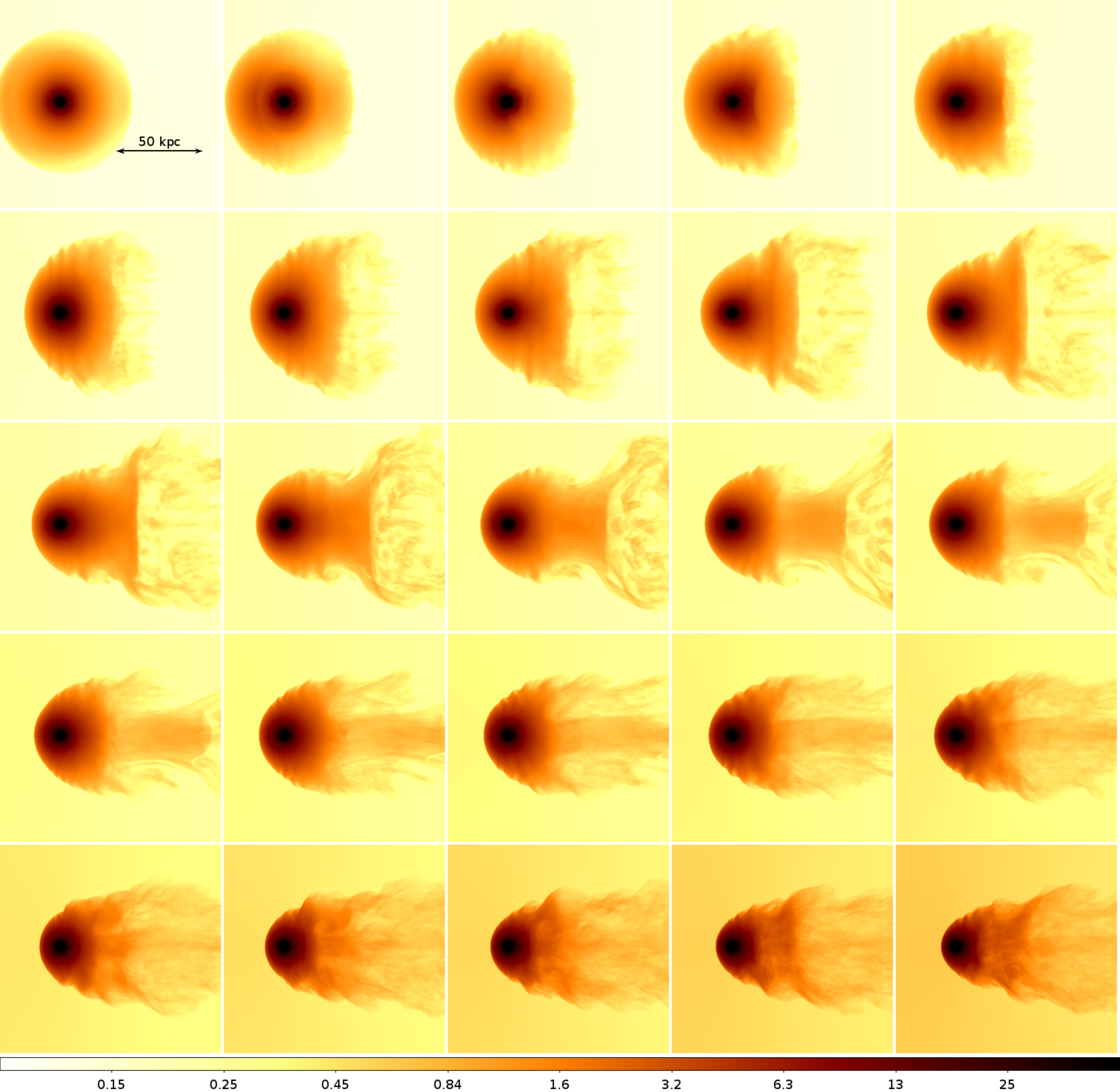}
\caption{Same as Fig.~\ref{fig:Xray_ini} but for  0.1 Spitzer viscosity.
\label{fig:Xray_ini_visc}}
\end{figure*}

Figures~\ref{fig:Xray_ini} to \ref{fig:Xray_ini_visc} display mock X-ray images from the relaxation phase (early infall) of inviscid and 0.1 Spitzer-viscous stripping of the initially extended atmosphere. The onset of the flow drives shocks through the atmosphere that are especially prominent in the first row in Fig.~\ref{fig:Xray_ini}. Such internal shocks could also be triggered if the galaxy encounters abrupt changes in ambient ICM flow due to  bulk motions in the cluster outskirts. 

In the following rows in Figs.~\ref{fig:Xray_ini} and \ref{fig:Xray_ini_visc}, the oscillation of the downstream atmosphere is clearly seen, leading to a variety of morphologies of the remnant atmosphere with and without a clear tail. Temporarily, the atmosphere is cone-shaped (most of second row of Fig.~\ref{fig:Xray_ini}) as seen in several  groups infalling into their host clusters or even in group mergers (e.g., in Abell 85, \citealt{Kempner2002}, or in RXJ0751, \citealt{Russell2014}). The cone shape is almost absent for the initially compact atmosphere, which is missing sufficient gas at large radii to form the cone of displaced galactic gas. For the extended atmosphere, the cone shape is preserved up to inclination angles of $45\degree$ (Fig.~\ref{fig:Xray_ini_rot}). For inclined galaxies, the KHIs along the sides appear less pronounced as they overlap and lead to a washed-out interface to the ICM. Similar to the remnant tail described above, the bright, unmixed gas in the cone is no indicator of suppressed mixing of stripped gas. Instead, the cone-shape arises from the deformation of the remnant atmosphere, the whole cone is still part of the remnant atmosphere. 

As discussed above, the global structure of the remnant atmosphere is independent of viscosity. However, in inviscid stripping KHI rolls are present along the sides of the remnant atmosphere. These are largely suppressed in the 0.1 Spitzer-viscous stripping. In the later stages in Fig.~\ref{fig:Xray_ini_visc}, stripped gas marks the wake of the galaxy. In the inviscid case, stripped gas in the wake is efficiently mixed and thus very faint.

\subsection{Quasi-steady stripping}


\begin{figure*}
\textbf{initially compact atmosphere}\\
inviscid stripping \hfill 0.1 Spitzer viscosity \hfill \phantom{x}\\
\includegraphics[trim= 110 0 400 0,clip,angle=0,width=0.44\textwidth]{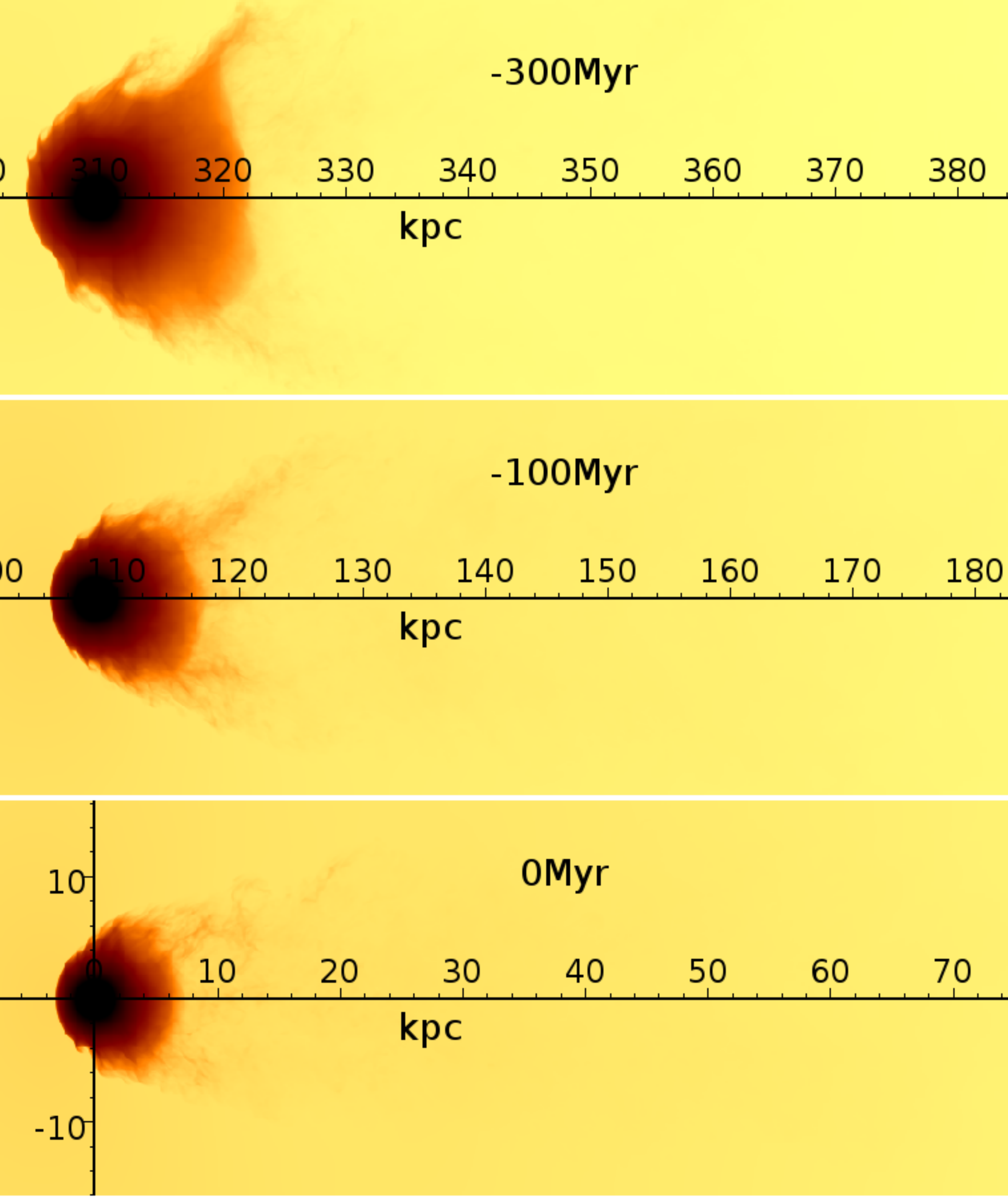}
\includegraphics[trim= 110 0 400 0,clip,angle=0,width=0.44\textwidth]{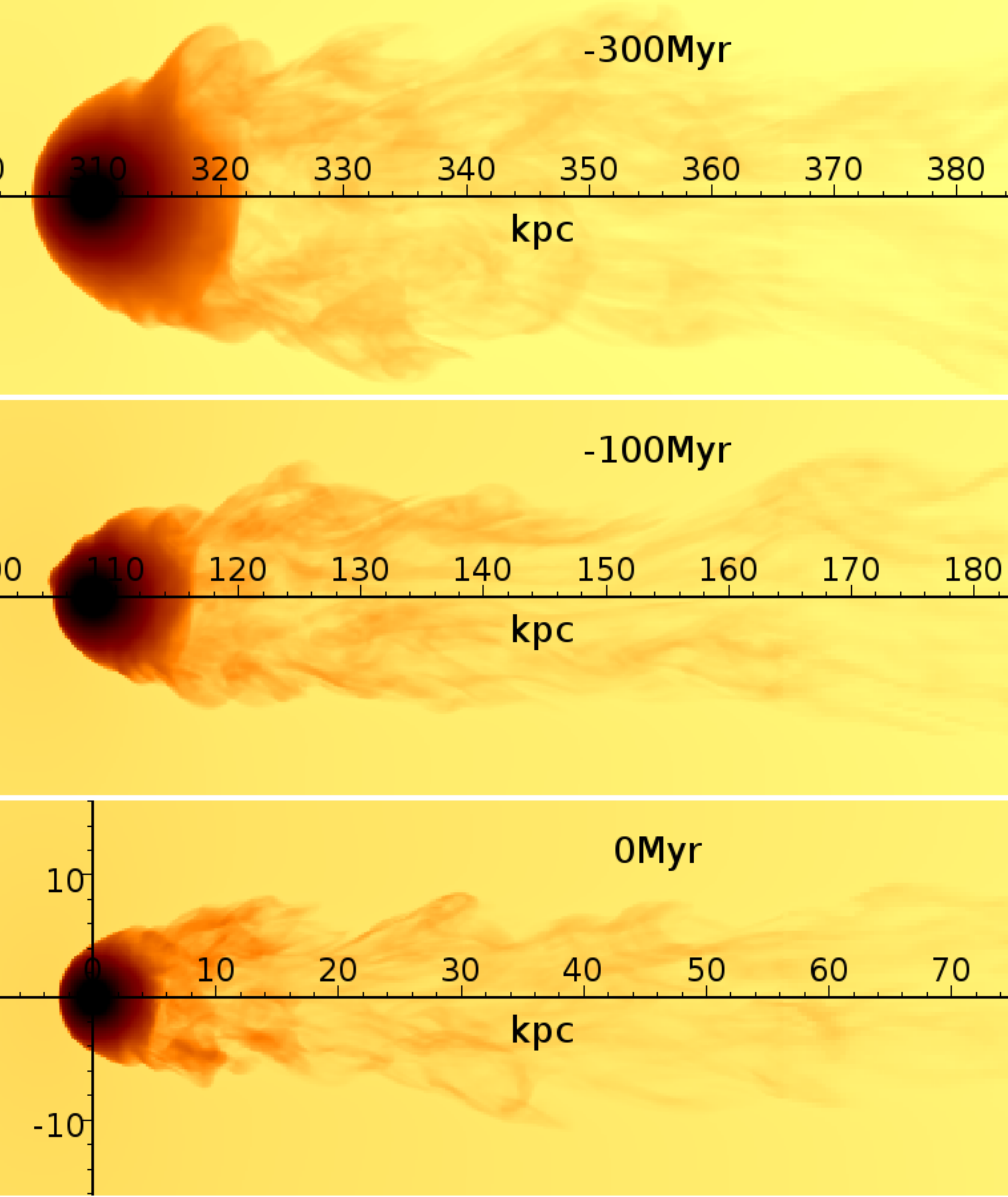}
\includegraphics[trim= 1100 0 0 0,clip,angle=0,width=0.095\textwidth]{CompSp350-500}
\caption{X-ray images (0.7-1.1 keV) of simulated gas-stripped galaxies, surface brightness in arbitrary units. Initially compact atmosphere, inviscid and 0.1 Spitzer viscosity in left and right column, respectively. Time steps as labelled, pericenter passage occurs at $t=0$.  In inviscid stripping, stripped gas in the wake is mixed with the ICM, leading to a very faint wake.
Viscosity suppresses KHIs and mixing in the wake, thus, unmixed cool dense gas makes the wake bright. 
\label{fig:Xray_comp}}
\end{figure*}

\begin{figure*}
\textbf{initially extended atmosphere}\\
inviscid stripping \hfill 0.1 Spitzer viscosity \hfill \phantom{x}\\
\includegraphics[trim= 70 0 370 0,clip,angle=0,width=0.43\textwidth]{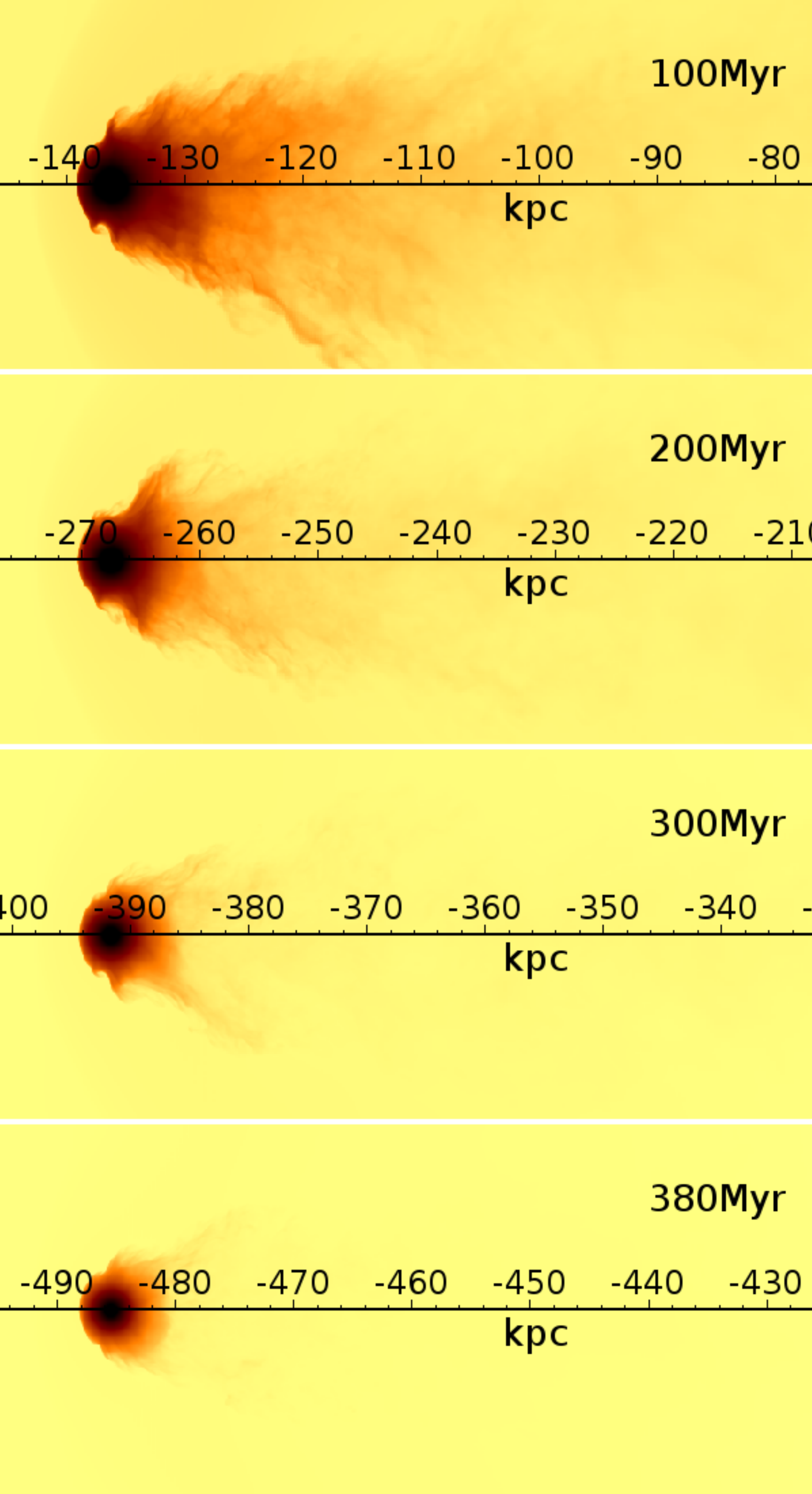}
\includegraphics[trim= 70 0 370 0,clip,angle=0,width=0.43\textwidth]{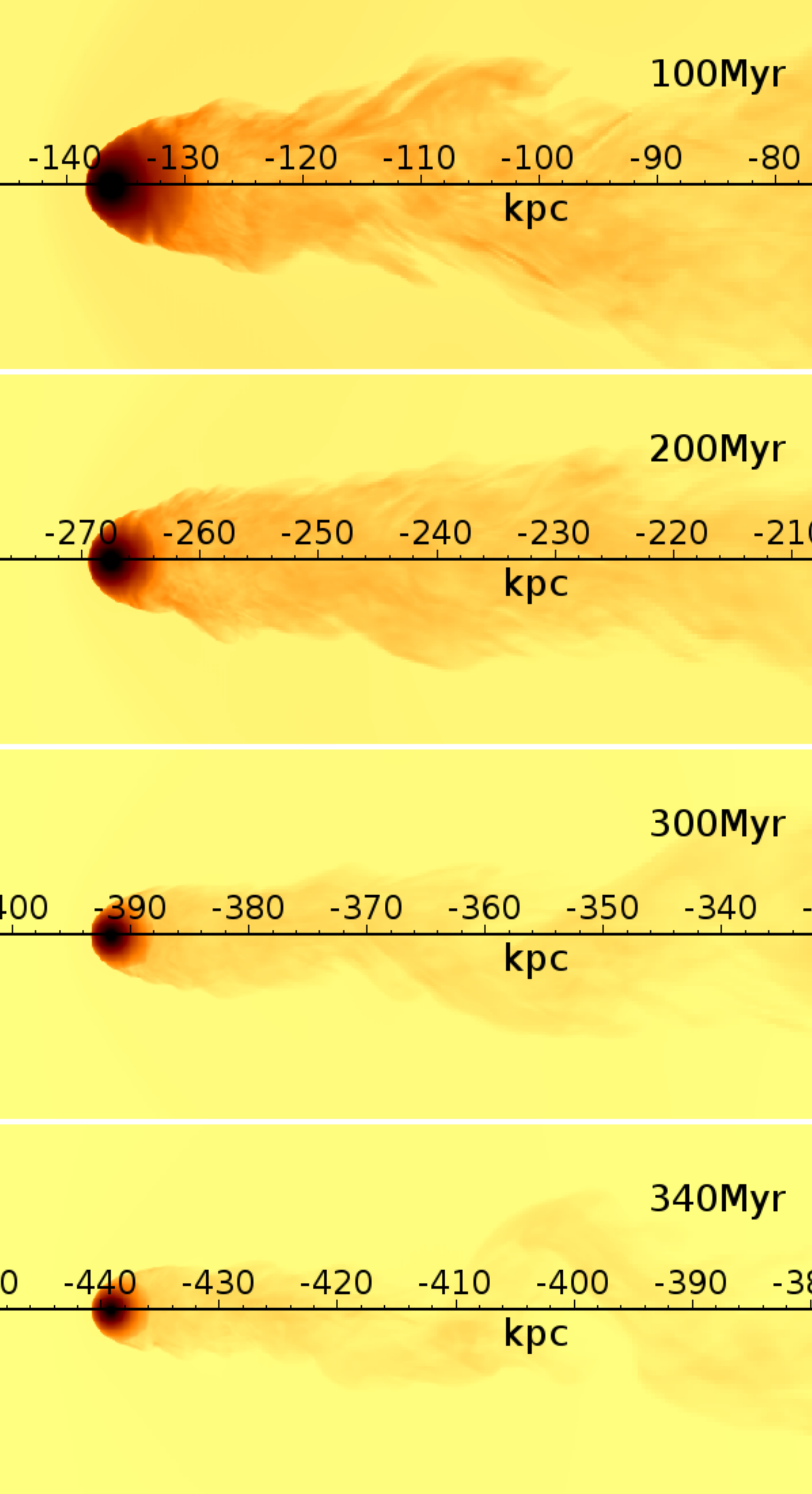}
\includegraphics[trim= 875 0 0 0,clip,angle=0,width=0.11\textwidth]{ExtVisc750-900}
\caption{Mock X-ray images (0.7-1.1 keV) of simulated gas-stripped galaxy, surface brightness in arbitrary units. Initially extended atmosphere, inviscid and 0.1 Spitzer viscosity in left and right column, respectively; quasi-steady stripping phase. Time after pericenter passage as labelled. Due to the initially extended atmosphere the galaxy has a pronounced remnant tail that is eroded only after $t=300\Myr$.  The  near wake is brighter than in Fig.~\ref{fig:Xray_comp}. In inviscid stripping, the wake fades quickly due to mixing, but it does not in viscous stripping.
\label{fig:Xray_ext}}
\end{figure*}

\begin{figure*}
\textbf{initially extended atmosphere, inviscid stripping}\\
\includegraphics[trim= 0 0 0 0,clip,angle=0,width=0.99\textwidth]{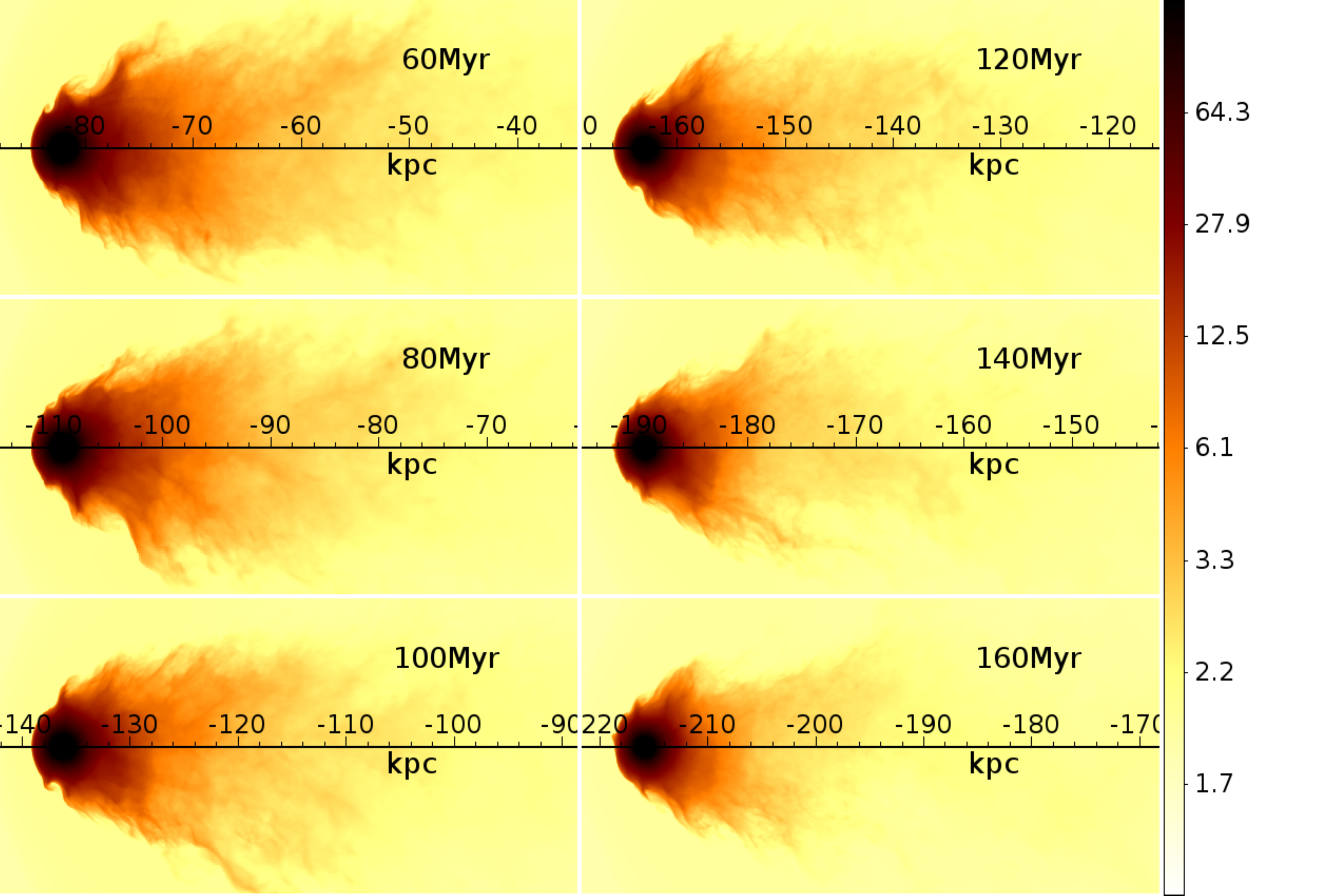}
\caption{Mock X-ray images (0.7-1.1 keV) of simulated gas-stripped galaxy, surface brightness in arbitrary units. Initially extended atmosphere, inviscid stripping. Time steps as labelled; for this epoch the tail brightness is in the same range as observed for M89 (paper III). The remnant tail has a smooth side edge, but the far tail is fuzzy and fades rapidly. The exact position and size of the KHIs varies quickly.
\label{fig:Xray_best}}
\end{figure*}

\begin{figure*}
\includegraphics[trim= 0 130 100 30,clip,angle=0,width=0.99\textwidth]{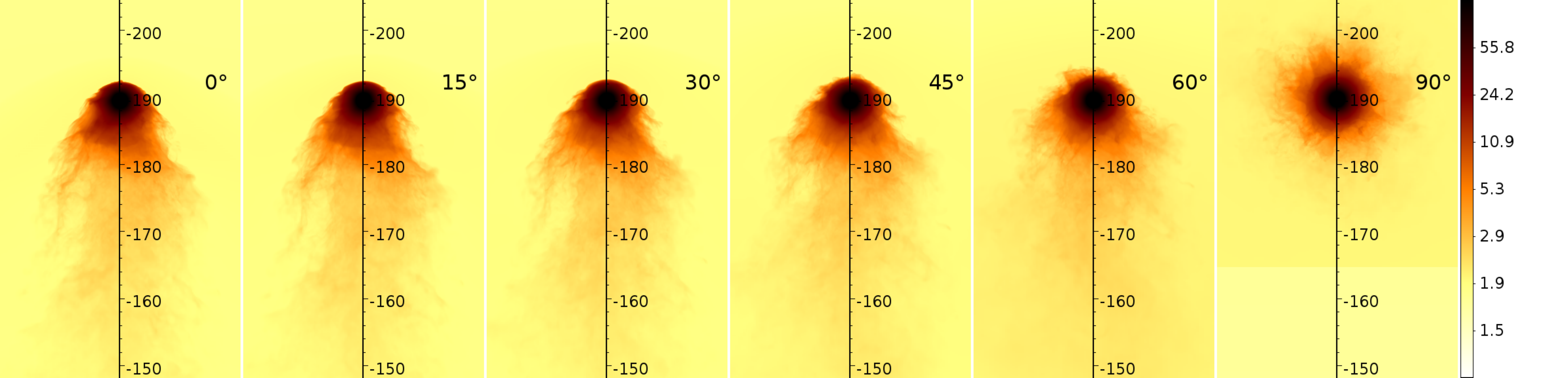}
\caption{Mock X-ray images (0.7-1.1 keV) of simulated gas-stripped galaxy, surface brightness in arbitrary units. Initially extended atmosphere, inviscid stripping, 140 Myr after pericenter passage, for different inclinations of the orbital plane (rotation around the horizontal axis, see labels in each panel). The axis is labelled in kpc and shows the distance to pericenter. Note the slightly bent appearance of the remnant tail. Paper III contains this figure with a different colorscale highlighting the projected position of the bow shock.
\label{fig:Xray_inclination}}
\end{figure*}

\begin{figure}
\textbf{initially extended atmosphere}\\
spatially constant kinematic viscosity\\
\rotatebox{90}{ \phantom{xxxx} 400 Myr \phantom{xxxxxxxxx} 200 Myr \phantom{xxxxxxxxxxx} 0 Myr}
\includegraphics[trim= 0 0 0 0,clip,angle=0,width=0.48\textwidth]{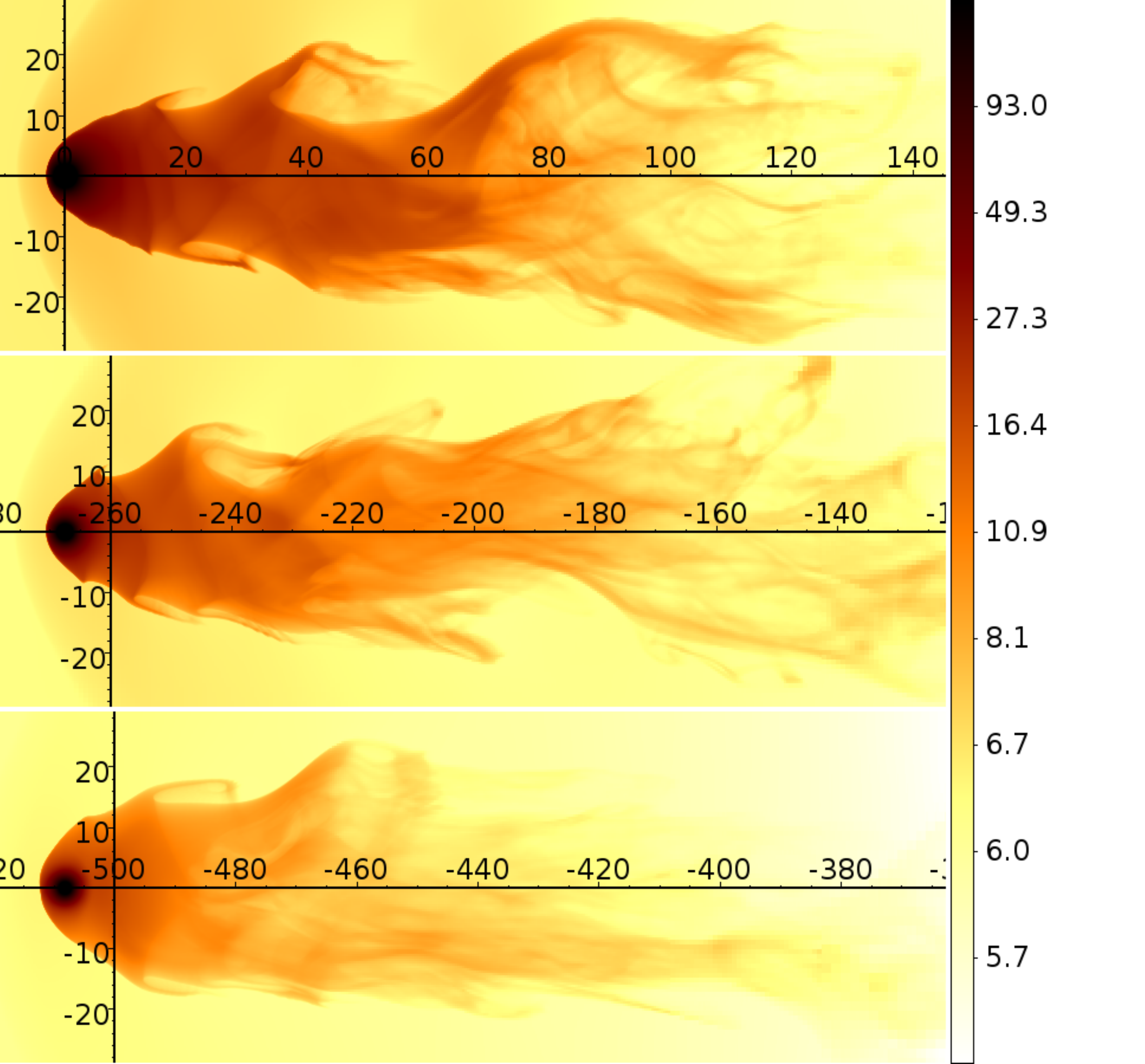}
\caption{Mock X-ray images of viscous stripping of the initially extended atmosphere with a spatially constant kinematic viscosity, i.e., also the galactic gas is viscous. Strongly suppressed mixing leads to an extremely long and bright remnant tail.
Note the length scale. Axes are labelled in kpc. 
\label{fig:Xray_constvisc}}
\end{figure}


Figures~\ref{fig:Xray_comp} and \ref{fig:Xray_ext} show  time series of X-ray snapshots for the inviscid and 0.1 Spitzer-viscous simulations around pericenter passage, for the compact and extended atmospheres, respectively. The X-ray emission traces most of the flow patterns described above:
\begin{itemize}
\item Prior to and near pericenter passage, the stripped galaxies display a head-tail structure. The downstream radius is larger than the upstream radius, which is the manifestation of the remnant tail. The remnant tail is eroded only after pericenter passage.
\item The brightness and  morphology of the  remnant tail  depends mostly on the initial gas contents. In case of the initially compact atmosphere (Fig.~\ref{fig:Xray_comp}) the remnant tail is not very prominent and the overall structure would be described as an asymmetric atmosphere. There is a clear downstream contact discontinuity between the remnant tail and the wake. For the initially extended atmosphere (Fig.~\ref{fig:Xray_ext}), the atmosphere has a long remnant tail (still $\sim 15\Kpc$ at pericenter passage) that is eroded only $\sim 300\Myr$ after pericenter passage. In projection, the remnant tail and wake are not clearly separated but combine into one tail-like structure. 
\item For the Spitzer-viscous cases, the wake decreases only slowly in brightness because the stripped gas  mixes only very slowly. For the initially compact atmosphere the wake has a filamentary structure;  for the initially extended atmosphere filaments in the wake are heavily superimposed, but give the wake a hatched appearance. 
\item In the inviscid cases, the brightness of the {\textit{wake} (beyond the remnant tail) and especially of the far wake} decreases faster than in the corresponding viscous cases. For the initially compact atmosphere, the wake is limb-brightened because the gas removal occurs only along the sides of the galaxy, and the wake is `hollow'. This effect is only mild for the initially extended atmosphere because stripped gas is also trapped in the inner part of the wake due to the deadwater region. 
\item Inviscid stripping leads to several horn-like or wing-like KHI rolls along the sides of the atmosphere and remnant tail. The horns or rolls start  near the upstream edge with a size smaller than the upstream radius, and grow along the sides of the atmosphere. At the end of the remnant tail they reach sizes larger than the current upstream radius, but comparable to the local radius of the tail. The exact position and shape of the horns varies rapidly with time as shown in Fig.~\ref{fig:Xray_best}.   The upstream side of the upstream-most rolls is generally sharp, but downstream sides of KHI rolls are diffuse and filled with faint emission. Further down the remnant tail the superposition of KHI rolls gives the remnant tail a diffuse boundary. Synthetic images taken along inclined LOSs confirm the filamentary nature of the KHI rolls (Fig.~\ref{fig:Xray_inclination}).
\item In contrast,  prominent horns are absent at 0.1 Spitzer viscosity near pericenter passage because this level of viscosity suppresses also the largest-possible KHIs for an atmosphere of this size. At earlier times, when the atmosphere is still larger, larger horns can exist (Fig.~\ref{fig:Xray_comp}, top right panel). At high viscosity, the wake has a sharp but ragged boundary for tens of kpc or many atmosphere diameters.
\item At intermediate viscosities, the effect of viscosity is subtle. Qualitatively, a higher viscosity leads to slower mixing along the wake. However, distinguishing, e.g., Reynolds numbers of 500 and 5000 would require a  detailed knowledge of the initial gas contents of the galaxy beyond what is possible. Alternatively, one could attempt to observe the absence or presence of KHI rolls below the length scale of the remnant atmosphere diameter, which, however, requires very high spatial resolution, and is  difficult due to projection effects. Thus, a stripped elliptical with a remnant atmosphere most easily probes the viscous suppression of KHIs on scales comparable to the atmosphere's radius.
\item In surface brightness images, the transition from the deadwater region to the far wake is hard to distinguish.
\end{itemize}

The X-ray images shown above assumed a LOS  perpendicular to the orbit of the galaxy. If the galaxy orbit is inclined out of the plane of the sky KHIs at the sides of the atmosphere start to overlap and lead to a somewhat fuzzier boundary of the atmosphere. From a certain inclination on (here $45\degree$) KHI rolls from the far side of the atmosphere appear at the upstream edge (Fig.~\ref{fig:Xray_inclination}). 

Fig.~\ref{fig:Xray_constvisc} displays X-ray snapshots of the stripping of the extended atmosphere with spatially constant viscosity. {Layers of gas are  being peeled off slowly along the remnant atmosphere and remnant tail,} forming  horns of  $\gtrsim 10 \Kpc$. The  downstream end of the remnant tail has a frazzled appearance. Note the large spatial scale of the remnant tail in this case. It certainly does not match M89 in terms of tail length and morphology.

\section{Discussion} \label{sec:discussion}
In the following subsections, we briefly compare our simulation results to our target galaxy M89, discuss possible effects of magnetic fields, and apply our results to further gas stripped galaxies.

\subsection{Comparison to M89} \label{sec:compare_m89}
Paper III of our series (Kraft et al.) presents a detailed comparison between our simulation results and new deep Chandra observations of M89. We show that inviscid stripping of our model galaxy with the extended atmosphere shortly after pericenter passage matches the observations best. We suggest that the bright cool near part of M89's tail is a remnant tail of still unstripped gas of the downstream atmosphere.  The sudden flaring and decrease in brightness of the far tail denotes the onset of the wake. The decrease in brightness of the wake, the absence of a bright wake in archival XMM data, and the presence of the KHI-like horns at M89's remnant atmosphere argue against a substantial isotropic viscosity in the Virgo ICM.

\subsection{Impact of magnetic fields}
{The comparison between M89 and our simulations disfavors only high \textit{isotropic} ICM viscosities. With anisotropic, i.e., Braginskii viscosity (\citealt{Kunz2012}), the shear layer along the galactic atmosphere may behave differently because magnetic fields in this shear layer could be largely aligned with the shear flow and the Braginskii viscosity could not suppress all KHI modes. If the aligned magnetic field itself is sufficiently weak, KHIs could grow despite Braginskii viscosity (see also \citealt{Suzuki2013}).}

Dense gas clouds moving through a magnetized plasma are predicted to be draped by the ambient magnetic fields (\citealt{Asai2007,Dursi2008}), at least in idealized field geometries. The magnetized layer wrapped around the gas cloud could suppress KHIs and mixing in its wake as well. Given that we do observe KHIs and mixing in the wake, draping may not be as efficient with tangled magnetic fields. \citet{Ruszkowski2007} showed this effect for buoyantly rising bubbles in the ICM. Only tangled fields with coherence lengths larger than the bubble can prevent its breakup by instabilities. If this analogy can be applied directly to galaxy stripping, the \textit{presence} of KHIs and mixing in the tail would indicate tangling of the Virgo magnetic fields on scales as small as $\sim 3\Kpc$.

\subsection{Speculations for other galaxies}

We apply our simulation results qualitatively to other stripped ellipticals and briefly discuss  their stripping state and implications for the ICM viscosity. 

M49 (NGC 4472) in the southern outskirts of the Virgo cluster has a ragged upstream (northern) edge, a wing in the east, and a flaring tail to the south-west (\citealt{Kraft2011}). At a distance of 1.4 Mpc from the Virgo center, M49 could still be in the flow initialization phase, which could explain the slight mismatch between the indicated directions of motion by the south-western tail and the northern edge. Due to its larger size and lower ambient ICM temperature, even at full Spitzer viscosity the Reynolds number of the ICM flow around M49 is Re $\sim 100$. Large-scale KHI may be present even at full Spitzer viscosity, but they should be very clear-cut because smaller-scale KHIs would still be suppressed. 

M86 (NGC 4406), the center of the subgroup 400 kpc west of the Virgo center, has a $\sim 150\Kpc$ long cool bright bifurcating `tail' that emerges not from M86 itself, but from a plume 20 kpc north of the galaxy (\citealt{Randall2008}). At first glance, this apparently unmixed tail could indicate a significant viscosity or draped magnetic fields. However,  H$\alpha$ and HI observations of the M86 group reveal bridges of ionized and neutral gas from the spiral galaxies  NGC 4438 and 4388 to M86 and beyond, respectively (\citealt{Kenney2008}, \citealt{Oosterloo2005}). These structures indicate possible encounters of these galaxies with M86 and thus a complex dynamical history of the M86 group, which could be responsible for the peculiar geometry of this tail and its apparently suppressed mixing. Disentangling this system's dynamic history will be the focus of a future paper.

The Fornax elliptical NGC 1404 is, in projection, 60 kpc  close to the Fornax cluster center (NGC 1399). It has a contact discontinuity towards the Fornax center and a faint `tail' on its opposite side (\citealt{Machacek2005}). We performed first gas-stripping simulations for this galaxy (Roediger et al., in prep.). In the compact Fornax cluster atmosphere the initial relaxation phase lasts up to pericenter passage.  Consequently, independent of viscosity, NGC 1404 should have a bright long remnant tail and a bright near wake if it was on its first infall into the Fornax cluster. The absence of either of these features, but presence of the faint `tail' supports the assumption that NGC 1404 has already lost its outer atmosphere on a first passage through Fornax. Its `tail' is thus its near wake. The same scenario was suggested by \citet{Bekki2003} because NGC 1404 appears to have lost a significant fraction of its globular clusters to NGC 1399 in the Fornax center. Due to the lower Fornax ICM temperature the Reynolds number of the ICM flow around NGC 1404 is $\sim 170$ at full Spitzer viscosity, which should be sufficient to suppress small-scale KHIs, but not the ones on scales of the atmosphere diameter. No obvious KHIs can be identified in the current data, though.

\section{Summary} \label{sec:summary}

We determined the impact of an isotropic, Spitzer-like viscosity on gas stripping of elliptical galaxies, in particular, for the Virgo elliptical M89 (NGC 4552). We showed that the evolution of the remaining galactic atmosphere proceeds similarly in Spitzer-viscous and in inviscid stripping. However,  10\% of the nominal Spitzer viscosity, or  a Reynolds number $\sim 50$ w.r.t.~the ICM flow around the remnant atmosphere, already leads to observable differences in the stripping process. We provide mock X-ray images for different stripping stages and conditions.

We show that the brightness of the near wake and the length of the galaxy's remnant tail depend on the galaxy's initial gas contents and are thus not suitable to determine ICM properties. In particular, the remnant tail cannot be used as a tracer for ICM transport properties because it is simply the remnant of the downstream atmosphere, hence it consists of unstripped and unmixed galactic gas even in inviscid stripping.  

Observationally, the effect of viscosity can be distinguished in two locations: 
\begin{itemize}
\item At the sides of the remnant atmosphere and its remnant tail, inviscid and high-Reynolds number stripping leads to ubiquitous KHIs that appear as horns or wings. These are absent with sufficient viscosity.
\item  Second, in Spitzer-like viscous stripping at a Reynolds number of $\sim 50$, suppressed mixing in the galaxy's \textit{wake} would allow the denser, cooler stripped galactic gas to survive in the wake, leading to a bright, cool \textit{wake} extending beyond 5 times the length of the remnant atmosphere. At high $\Reyn$, mixing of the stripped galactic gas quickly reduces the gas density and thus X-ray brightness in the wake. 
\end{itemize}

A detailed comparison to archival XMM and new Chandra data is presented in  paper III of this series (Kraft et al., in preparation).

\acknowledgments
The FLASH code was in part developed by the DOE NNSA- ASC OASCR Flash center at the University of Chicago. E.R.~acknowledges the support of the Priority Programme Physics of the ISM of the DFG (German Research Foundation),  the supercomputing grants NIC 6006 and 6970 at the John-Neumann Institut at the Forschungszentrum J\"ulich,  a visiting scientist fellowship of the Smithsonian Astrophysical Observatory, and the hospitality of the Center for Astrophysics in Cambridge. We also thank the referee for helpful comments regarding the presentation of our results.



{\it Facilities:} \facility{CXO (ACIS)}.

%
\bibliographystyle{apj}
\bibliography{library}


\appendix

\section{Flow around a solid obstacle} \label{sec:solidbodysims}


\begin{figure*}
\begin{sideways}
\begin{minipage}{\textheight}
\includegraphics[angle=0,width=0.99\textheight]{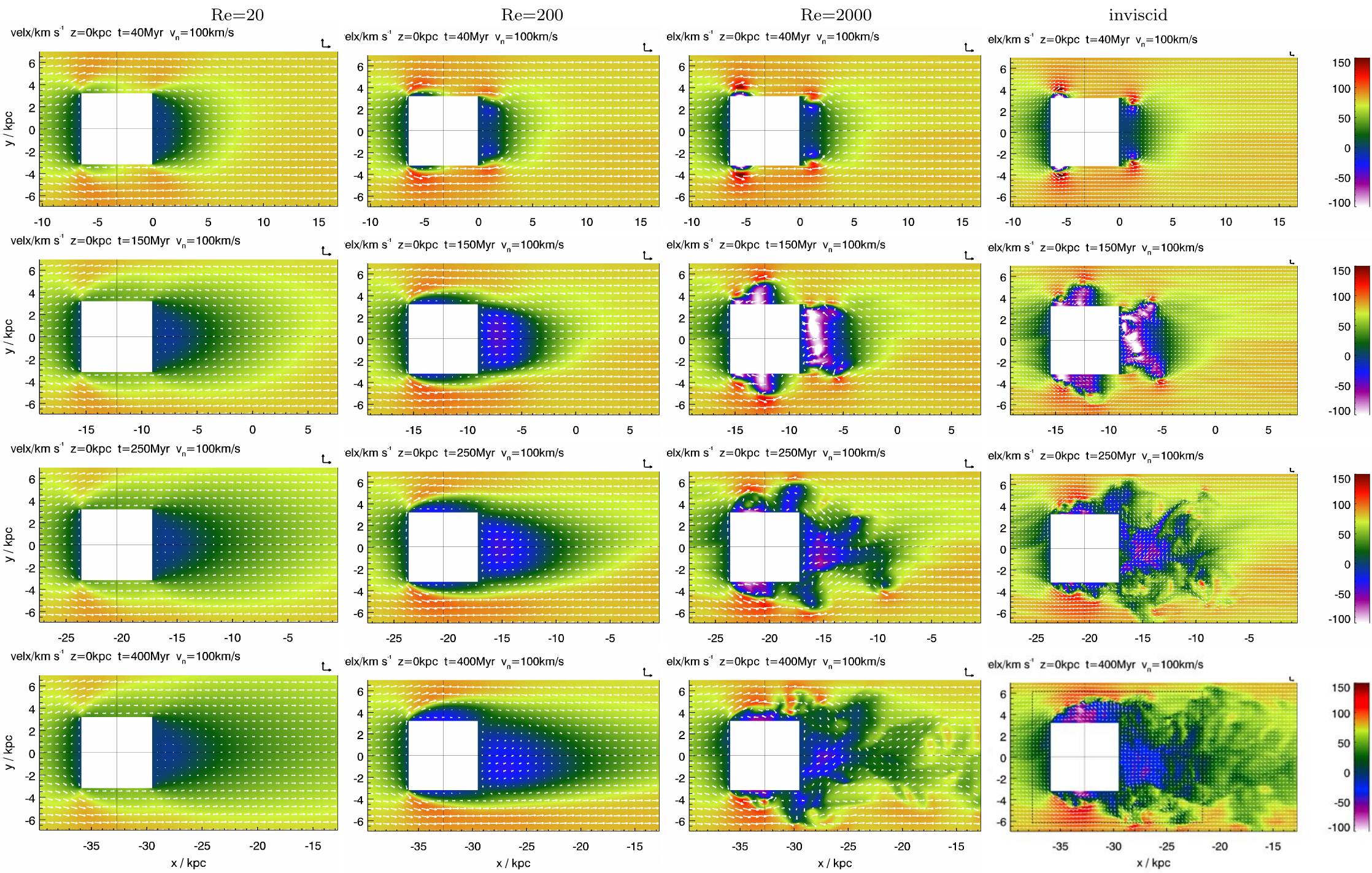}
\caption{Flow around a solid cube at different Reynolds numbers (columns) and timesteps (rows), simulated with FLASH. The color codes the $x$-component of the velocity $v_x$,   velocity vectors are overlaid. The flow was initialized at $t=0$ with 80$\Kms=$Mach 0.1. The flow is perturbed three times at $t=0,100,200\Myr$ by adding a superposition of sin-waves to $v_x$ with an amplitude of 8$\Kms$ each, as evident in the early inviscid panels. In the inviscid case, it takes about 4 crossing times (cube size/flow velocity) for the typical flow pattern to be established. The simulations use a nested grid. Around the cube and in the wake (marked by black dashed box in the bottom left panel) the resolution is 0.05 kpc (128 cells per cube length). Beyond that, the resolution is 0.1 kpc throughout the shown portion of the grid. The total simulation volume is $(-30\Kpc, 60\Kpc)\times(\pm 30\Kpc)^2$.}
\label{fig:boxflow}
\end{minipage}
\end{sideways}
\end{figure*}

\begin{figure*}
\hfill Inviscid, $dx=1/128$th cube length \hfill Inviscid, $dx=1/64$th cube length \hfill $\Reyn=1000$, $dx=1/128$th cube length \hfill\phantom{x}\newline
\includegraphics[angle=0,width=\textwidth]{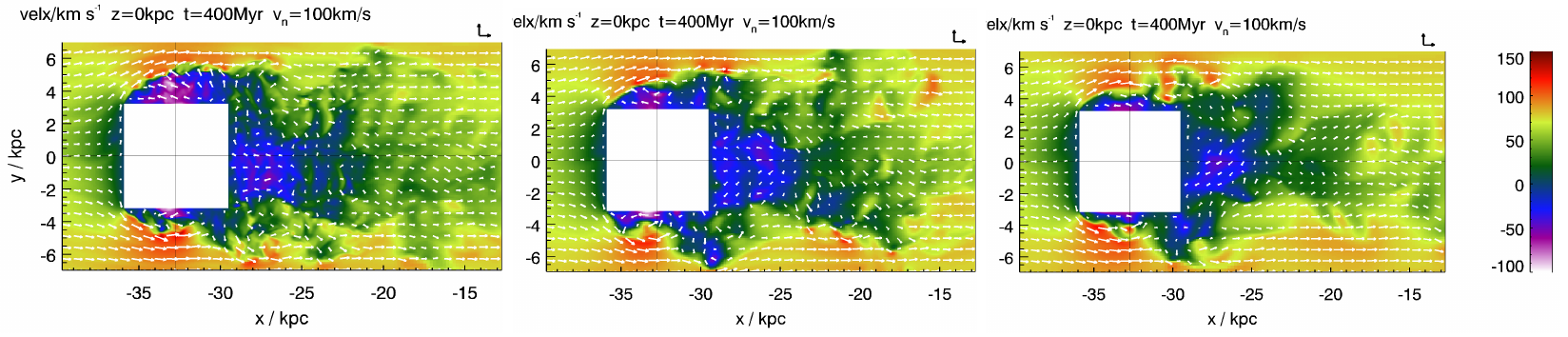}
\caption{Flow past a cube, comparison of inviscid simulation at resolution 1/128 per cube length and 1/64 per cube length, and with viscous simulation with $\Reyn=1000$ at the higher resolution.}
\label{fig:box_res}
\end{figure*}

We tested the ability of the FLASH code to capture flows past solid bodies at different Reynolds numbers. The simulation grid in FLASH is organized in blocks, in our case of size of $16^3$ grid cells per block.  FLASH4 provides the option to apply reflecting boundary conditions to one or more blocks of the lowest refinement level inside the simulation domain, i.e., to treat them as solid obstacles. We utilized this option to simulate the flow past a solid cube as shown in Fig.~\ref{fig:boxflow}. We chose a cube size, wind density and temperature of $6.4\Kpc$, $10^{-27}\gccm$ and of 2.4 keV, respectively, i.e., comparable to the atmosphere of M89, but of course the results can be scaled by Reynolds number and  Mach number. The wind velocity is subsonic at Mach 0.1 ($80\Kms$). With these length and velocity scales, the kinematic viscosities of $8\times 10^{27}$, $8\times 10^{26}$ and $8\times 10^{25}$ cm$^2$s$^{-1}$ correspond to Reynolds numbers of 20, 200, and 2000.  We also ran an inviscid reference simulation. We use a nested grid with a peak resolution of 1/128th of the cube length around the cube and in the near wake as indicated in Fig.~\ref{fig:boxflow}. 

With increasing Reynolds numbers, the flow patterns change as expected for the flow past a 3D object from a vortex-free, laminar flow to a downstream vortex ring (\citealt{Taneda1956}) to an irregular near wake. At our resolution, the inviscid flow shows smaller-scale structures than the $\Reyn=2000$ flow, i.e., the numerical viscosity is smaller than the physical viscosity at $\Reyn=2000$. Fig.~\ref{fig:box_res} demonstrates the effect of decreasing resolution and thus increasing numerical viscosity.

\section{Resolution test}  \label{sec:resolution}


\begin{figure*}
inviscid \hfill 0.1 Spitzer viscosity \hfill\phantom{x}\\
\rotatebox{90}{$\Delta x=0.4\Kpc$\hspace{3cm} $\Delta x=0.1\Kpc$}\includegraphics[trim= 0 0 0 0,clip,angle=0,width=0.41\textwidth,origin=c]{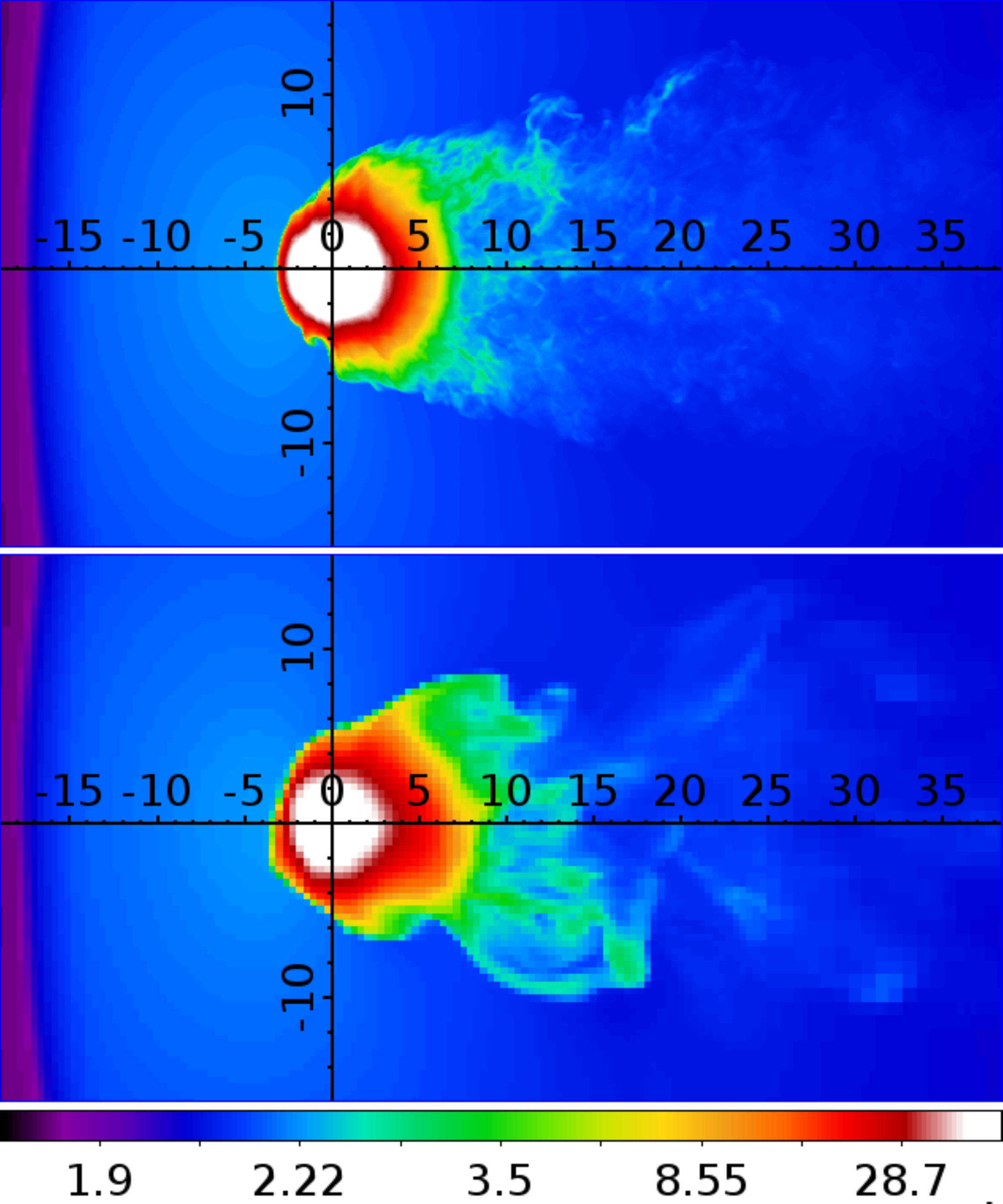}
\hfill
\includegraphics[trim= 0 0 0 0,clip,angle=0,width=0.5\textwidth,origin=c]{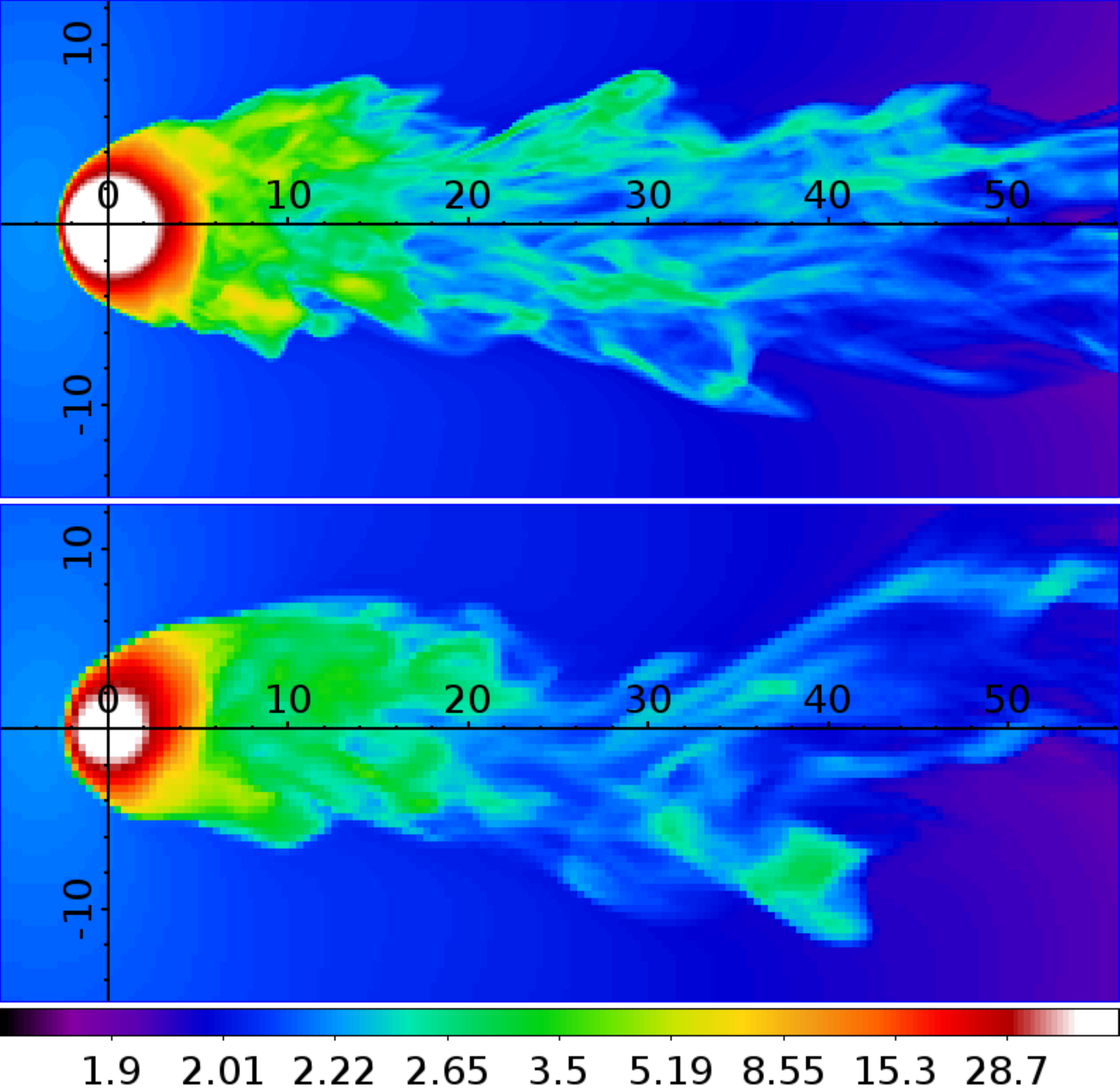}
\rotatebox{90}{$\Delta x=0.4\Kpc$\hspace{3cm} $\Delta x=0.2\Kpc$}
\caption{Resolution test: X-ray images (0.7-1.1 keV) of simulated gas-stripped galaxy, surface brightness in arbitrary units. The color stretch highlights the tail. Initially compact atmosphere, inviscid (left) and viscid (0.1 Spitzer viscosity, right) stripping, at pericenter passage. The bottom panels show result for a peak spatial resolution of 0.4 kpc, the top panels are for resolutions of 0.1 kpc in case of inviscid stripping and 0.2 kpc in case of viscid stripping.
\label{fig:Xray_resolution}}
\end{figure*}

We repeated our galaxy stripping simulations at a factor of 2 or 4 lower resolution and find that our results converged regarding tail structure and brightness. Fig.~\ref{fig:Xray_resolution} shows example synthetic X-ray images for two tests.

\end{document}